\pdfoutput=1 

\documentclass[11pt,a4paper]{article}

\usepackage[
  a4paper,
  left=2.3cm,
  right=2.3cm,
  top=2.5cm,
  bottom=2.7cm
]{geometry}

\usepackage[T1]{fontenc}
\usepackage[utf8]{inputenc} 
\usepackage[english]{babel} 
\usepackage{lmodern}
\usepackage{microtype}

\usepackage{amsmath,amssymb,amsfonts}
\usepackage{bm}        
\usepackage{mathtools}
\usepackage{siunitx}

\usepackage{bbold}

\usepackage{graphicx}
\usepackage{booktabs}
\usepackage{tabularx}
\usepackage{xcolor}
\usepackage{subcaption}

\usepackage[
  colorlinks=true,
  linkcolor=blue,
  citecolor=blue,
  urlcolor=blue,
  pdfauthor={First Author and Co-authors},
  pdftitle={Title of the manuscript}
]{hyperref}

\usepackage[nameinlink,capitalise,noabbrev]{cleveref}

\usepackage{authblk}

\title{\Large Floquet Theory and Average Hamiltonian Theory Revisited:\\ Equivalence, Convergence and Applications to NMR}

\author[1]{Antonia J. Bock\thanks{\href{mailto:antoniajoelle.bock@tu-dortmund.de}{antoniajoelle.bock@tu-dortmund.de}}}
\author[2]{Matthias Ernst\thanks{\href{mailto:maer@ethz.ch}{maer@ethz.ch}}}
\author[1]{Götz S. Uhrig\thanks{\href{mailto:goetz.uhrig@tu-dortmund.de}{goetz.uhrig@tu-dortmund.de}}}

\affil[1]{\scriptsize Department of Physics, TU Dortmund University,  Otto-Hahn-Str. 4, 44227 Dortmund, Germany}
\affil[2]{\scriptsize Department of Chemistry and Applied Life Sciences, ETH Zurich, Vladimir-Prelog-Weg 2, 8093 Zurich, Switzerland}

\date{} 

\begin{document}

\maketitle

\begin{abstract}
An accurate theoretical treatment of periodically driven quantum systems is crucial for various fields in 
the exact sciences, for instance Nuclear Magnetic Resonance (NMR) spectroscopy. 
Conventionally, either average Hamiltonian theory or Floquet theory 
is used to predict or describe experimental outcomes,
such as the time evolution or the spectra 
yielding the information of the sample under study. 
A detailed analysis of the equivalence of these two approaches 
with an emphasis on applications in NMR 
will help to improve the theoretical understanding of NMR experiments. 

In this work, we identify the Floquet--Magnus expansion as essential to prove the mathematical equivalence 
of Floquet theory and average Hamiltonian theory. 
We advocate a calculation scheme which is less prone to algebraic mistakes because explicit integration 
is avoided.  On this basis, we provide the first four orders of both theories. 
We further examine their applicability to some experiments in NMR. 
As examples, we investigate the Bloch--Siegert shift 
and dipolar coupled spin systems under magic-angle spinning. 

Based on our analysis, we recommend the use of the Floquet--Van Vleck approach 
including both the effective Hamiltonian and the kick operator. 
The consistent separation of secular and non-secular contributions 
appears to be especially advantageous for numerical robustness.  
Its accuracy is about three times better than the one provided by average Hamiltonian theory
despite their formal equivalence. 

Our findings provide important insights into 
the theoretical background of Floquet theory and average Hamiltonian theory.
This includes the extent of their algebraic and perturbative equivalence, 
with an emphasis on how these findings are of relevance in the analysis 
of solid-state NMR experiments. 

\end{abstract}

\begin{center}
\small\textit{This manuscript has been submitted to \emph{Magnetic Resonance}.}
\end{center}


\section{Introduction}

Periodically driven systems play a central role in a wide range of physical applications, from Floquet engineering \cite{Goldman2014,oka2019}, quantum control and sensing \cite{Zhou2020,oonAverageHamiltonianTheory2026} 
to ultracold atoms in driven optical lattices \cite{eckardtSuperfluidInsulatorTransitionPeriodically2005,Dengis2025},  
radiofrequency pulse sequences \cite{Brinkmann2004,PILEIO200765,Hou2012}, 
and sample spinning \cite{andrewNuclearMagneticResonance1958,andrewRemovalDipolarBroadening1959,loweFreeInductionDecays1959} in solid-state Nuclear Magnetic Resonance (NMR) experiments.
In quantum mechanics, the dynamics of these systems is governed by the time-dependent Schrödinger equation, i.e., a first-order initial value problem, which arises across various disciplines \cite{blanesMagnusExpansionIts2009,arnalExponentialPerturbativeExpansions2020}. 
For a time-periodic Hamiltonian $H(t)=H(t+T)$, the exact formal solution can be written with Dyson's formula \cite{Dyson1949} as a time-ordered exponential, in which the time-ordering operator takes account for the non-commutativity of the Hamiltonian at different time points, $[H(t),H(t')]\neq 0$ for $t\neq t'$. 

Despite the fact that this formal solution is well-known, an analytical integration is often not possible, and approximate methods are applied.
In most cases, they exploit the presence of distinct time and energy scales, such as the energies in $H(t)$, the modulation frequency $\omega \sim 1/T$, or the length of the timescale $t_0\rightarrow t$ covering the dynamics of interest. 
Throughout this work, we use Hamiltonians in angular-frequency units, such that the propagator is defined by
\begin{equation}\label{eqn:schroedinger}
    \frac{d}{dt}U(t,t_0) = -iH(t) U(t,t_0)\,,\quad 
    U(t_0,t_0) = \mathbb{1} \,. 
\end{equation}

In the high-frequency regime $\|H\|\cdot T \ll 1$, i.e., when the driving frequency is much larger than the energies found in $H(t)$, different perturbative approaches can be used to approximate the dynamics. 
Two of the most common approaches are Average Hamiltonian Theory (AHT) \cite{magnus1954,wilcoxExponentialOperatorsParameter1967,EVANS196872,Haeberlen1968,haeberlenHighResolutionNMR2012,mehringPrinciplesHighResolution2012,blanesMagnusExpansionIts2009,brinkmannIntroductionAverageHamiltonian2016a} 
and Floquet-based methods \cite{floquetEquationsDifferentiellesLineaires1883,shirleySolutionSchrodingerEquation1965,leskesFloquetTheorySolidstate2010a,scholzOperatorbasedFloquetTheory2010a,IVANOV202117}, also known as coherent averaging and secular averaging, respectively.
Both aim to approximate the same dynamics, yet appear to lead to seemingly different or even conflicting results \cite{leskesFloquetTheorySolidstate2010a,mehringPrinciplesHighResolution2012}. 
This raises the fundamental question of how equivalent these approaches are, under which conditions one is preferable over the other, how differences in their algebraic structure influence the accuracy of their results, and whether a clear guideline for selecting the appropriate framework can be established.

In solid-state NMR, periodically driven Hamiltonians naturally arise, for example, from radio-frequency irradiation, repetitive pulse sequences, and sample spinning.
Here, AHT has become a standard tool for describing effective spin dynamics, while Floquet-based methods provide an alternative formulation that is particularly suited to the treatment of explicitly periodic Hamiltonians and sideband structures or Hamiltonians with multiple incommensurate time dependencies.
While it is known that both theories can be understood within a common mathematical framework – namely, the Floquet–Magnus Expansion (FME) \cite{casasFloquetTheoryExponential2001,blanesMagnusExpansionIts2009,manangaIntroductionFloquetMagnusExpansion2011,eckardtHighfrequencyApproximationPeriodically2015,mikamiBrillouinWignerTheoryHighfrequency2016} – this connection is often not yet fully appreciated in practice in the NMR community. 
The situation is further complicated by the rapidly increasing algebraic complexity of higher orders, which makes the resulting expressions prone to errors and can obscure their formal equivalence.
As a consequence, important aspects remain unclear, including the precise origin of discrepancies between the methods, the role of secular and non-secular terms, and the effect of the rotation in Floquet space on results in Hilbert space. 

In this work, we revisit AHT and Floquet theory from a unified perspective.
We demonstrate that both AHT and the Floquet–Van Vleck (FVV) formalism arise as specific representations within the more general FME framework.
In doing so, the reader will be able to remove the shroud of apparent contradictions between AHT and FVV. 
In the following course, we identify and correct an inconsistency in the existing literature and introduce a computational scheme that avoids explicit time integration, thereby reducing the likelihood of errors. 
Based on this scheme, we derive the first four orders of both expansions in a form that enables the treatment of stroboscopic and non-stroboscopic measurements, allowing for the calculation of broader spectra with additional sidebands.
Finally, we assess the practical performance and numerical robustness of both approaches in the context of NMR spectroscopy by explicitly calculating and comparing spectra for representative use cases in NMR.

Our results show that FVV and AHT are consistent up to the approximate order in $1/\omega$, however, FVV is consistently more accurate than AHT by a factor of $\simeq 3$.
Additionally, it becomes clear that non-secular terms are not per se negligible apart from potential special cases in NMR in which they vanish. 
Overall, this work clarifies the relationship between AHT and FVV, explains the origin of their differences, and provides concrete guidelines for their applications in NMR.

The paper is structured with sections on the following topics.
We review the relevant literature on AHT, FVV, and how both are unified by the more general FME in Sect.~\ref{sec:theory}. 
This is followed by Sect.~\ref{sec:orders}, in which we provide the first four orders of both theories for time-periodic Hamiltonians, including a recipe of how they can be derived without integration in the frequency space in App.~\ref{app:calc_scheme}.
This section is also motivated by a typo that we have found in literature on our way, and we intend to provide consistent results in a better-suited notation that we propose.
In conclusion, we present the results of our numerical validation, which verify our formulas.
In Sect.~\ref{sec:NMR_appli}, we compare both approaches in typical NMR applications, demonstrating the advantages of FVV and the influence of the Van Vleck rotation in Floquet theory.
We finalize this work in Sect.~\ref{sec:summary} with a summary, outlook and discussion.

\section{Theory}
\label{sec:theory}

From a historical point of view, Average Hamiltonian Theory (AHT) and Floquet theory follow derivations that are 
independent of each other. 
Based on the introduction of Floquet theory subjected to first-order periodic linear differential equations in 1883 
\cite{floquetEquationsDifferentiellesLineaires1883}, 
it was initially extended to NMR to solve the time-dependent Schrödinger equation in 1965
\cite{shirleySolutionSchrodingerEquation1965}. 
Originally, a finite-dimensional truncation of the infinite-dimensional matrix representation \cite{BaldusLevanteMeier+1994+80+88,Levante10121995}
or a perturbative treatment of the matrix representation \cite{SchmidtVega1992} 
was used to allow numerical calculations and analytical understanding. 
For a better analytical understanding, block-diagonalization in the infinite-dimensional so-called Floquet space can be implemented in order to determine an effective Hamiltonian in the spin-Hilbert space. 
This can be done in various ways. A standard approach is perturbative in nature, the  Van Vleck perturbation theory \cite{vanvlecksigmaTypeDoublingElectron1929}, 
shown by Shimon Vega in 1996 \cite{Vega1996}. 
In this work, the combination of Floquet theory and Van Vleck perturbation theory is denoted 
the Floquet--Van Vleck approach (FVV). 

Parallel to this development, 
Wilhelm Magnus formulated the Magnus expansion in 1954 \cite{magnus1954} addressing purely algebraic issues 
arising in the context of time-dependent Hamiltonians. 
He derived a framework that enabled the calculation of a time-independent effective replacement for the Hamiltonian 
that describes the dynamics averaged over a sufficiently small time window. 
This method found various applications in the field of NMR. 
In 1968, Ulrich Haeberlen and John S. Waugh made use of the fact that for time-periodic Hamiltonians 
$H(t)=H(t+T)$ the state of the system at any integer multiple of the period $T$ is given 
once the evolution operator over a single cycle is known \cite{haeberlenCoherentAveragingEffects1968,haeberlenHighResolutionNMR2012}. 
This basically lays the foundation for the Average Hamiltonian Theory (AHT) 
which is the physical implementation of the Magnus expansion calculating the effective averaged single-cycle 
Hamiltonian in periodically modulated systems 
allowing to compute the state at the so-called stroboscopic time points $t=0,T,2T,3T,\dots$ 

Despite clear distinctions in their approach, AHT and FVV are 
found to be unified by a more general framework, typically referred to as 
the Floquet--Magnus Expansion (FME)
\cite{casasFloquetTheoryExponential2001,manangaIntroductionFloquetMagnusExpansion2011,eckardtHighfrequencyApproximationPeriodically2015,bukovUniversalHighFrequencyBehavior2015}. 
The original derivations of AHT and FVV might be more helpful in deepening the intuition of the corresponding theories, 
and various articles can be found in the existing literature, e.g., Refs.~\cite{blanesMagnusExpansionIts2009,leskesFloquetTheorySolidstate2010a,scholzOperatorbasedFloquetTheory2010a,IVANOV202117}. 
The derivation via the  FME bears the advantage that AHT and FVV can be understood in a unified picture
despite their distinct underlying ideas. 
Moreover, the formal equivalence between AHT and FVV is obvious in FME, 
and higher orders are more easily accessible. 
Thus, we first introduce FME and then show how to derive both AHT and FVV from it. 
For completeness, we include short sketches of the original historical idea of AHT and FVV, respectively.

\subsection{Floquet--Magnus expansion}

Starting point of the FME \cite{manangaIntroductionFloquetMagnusExpansion2011} 
is the Schrödinger equation, Eq.~(\ref{eqn:schroedinger}), of a time-periodic Hamiltonian. As it is customary in NMR, we write the Hamiltonian in angular frequency units
\begin{equation}
    H(t) = \sum_n H_n e^{in\omega t}\,,\quad \omega = 2\pi \nu = \frac{2\pi}{T}\,,
    \label{eqn:fourier_H}
\end{equation}
which is formally equivalent to applying 
natural units by $\hbar \equiv 1$.
For such systems, the Floquet theorem \cite{floquetEquationsDifferentiellesLineaires1883} 
states that the propagator splits up into a phase evolution 
governed by the time-independent, so-called effective Hamiltonian $F$
and a time-dependent periodic phase factor, the kick or micromotion operator $\Lambda(t)=\Lambda(t+T)$, 
describing the recurring dynamics within each period 
\begin{equation}
    U(t,t_0) = e^{-i \Lambda(t)} e^{-i F \cdot (t-t_0)} e^{i \Lambda(t_0)} .
    \label{eqn:floquettheorem}
\end{equation}
Because $H(t)$ is Hermitian, leading to the unitary nature of $U(t,t_0)$, 
both $F$ and $\Lambda$ are Hermitian. 
They are not unique and depend on the initial value $\Lambda(t_0)$, which is not fixed. 
The evolution over a single cycle is governed by the stroboscopic Hamiltonian~$\tilde{F}:=  e^{-i \Lambda(t_0)} F e^{i \Lambda(t_0)}$, i.e., 
\begin{equation}
    U(t_0+T,t_0) = e^{-i \Lambda(t_0)} e^{-i F \cdot T} e^{i \Lambda(t_0)} = e^{-i \tilde{F} \cdot T}\,. 
    \label{eqn:stroboscopicHamiltonian}
\end{equation}
The Floquet theorem can be viewed as the temporal analog to the Bloch theorem used for discrete 
translational symmetries in space. 
Analogously to Eq.~(\ref{eqn:fourier_H}), the Fourier series of the kick operator is defined by 
\begin{equation}
    \Lambda(t) = \sum_n \Lambda_n e^{in\omega t}\,.
    \label{eqn:LambdaFourier}
\end{equation}
The Fourier coefficients are calculated accordingly via 
\begin{subequations}
    \begin{align}
        H_n &= \frac{1}{T} \int_{0}^{T} \text{d}t\, H(t) e^{-in\omega t}\,, \\
        \Lambda_n &= \frac{1}{T} \int_{0}^{T} \text{d}t\, \Lambda(t) e^{-in\omega t}\,.
    \end{align}
\end{subequations}

Moreover, we define the $n$-times nested commutator 
\begin{align}
    [X,Y]_0 &:= Y \,, \,\,
    [X,Y]_{n\geq 1} := \bigl[X, [X,Y]_{n-1}\bigr]  \,. 
\end{align}
Note that this is the same as the Liouvillian commutator superoperator ${\cal L}_X$ and the adjoint operator $\text{ad}_X$, which are also used in the literature. 
This means $[X,Y]_n={\cal L}_X^n Y=\text{ad}_X^nY$.

The first step of FME is to insert Eq.~(\ref{eqn:floquettheorem}) into the Schrödinger equation, 
arriving at a differential equation of first order 
in the kick operator $\Lambda(t)$. 
Then it can be written in shorthand
\begin{subequations}\label{eqn:FME_dgl}
\begin{align}
    \frac{d\Lambda}{dt} 
    &= \sum_{k=0}^\infty \frac{B_k}{k!} (-i)^k {\cal L}_\Lambda^k\bigl\{ H + (-1)^{k+1} F \bigr\} \\ 
    &= \sum_{k=0}^\infty \frac{B_k}{k!}  \bigl([-i\Lambda, H]_k - [i\Lambda, F ]_k\bigr) \,. 
    \end{align}
\end{subequations}
Here, $B_n$ are the signed Bernoulli numbers \cite{AbramowitzSegun}, from which the first ones are $B_0=1$, $B_1=-1/2$, $B_2=1/6$, $B_3=0$, $B_4=-1/30$. 
By making use of their exponential generating function 
\begin{equation}
    f(x):= \frac{x}{e^x-1} = \sum_{n=0}^{\infty} \frac{B_nx^n}{n!}\,, 
    \label{eqn:BernoulliGenerating}
\end{equation}
Eq.~(\ref{eqn:FME_dgl}) is often rewritten as 
\begin{equation}
    \frac{d\Lambda}{dt} = f({\cal L}_{-i\Lambda}) H(t) - f({\cal L}_{i\Lambda}) F \,. 
\end{equation}
Both $\Lambda(t)$ and $F$ are expressed by perturbation series. 
The perturbation parameter can be viewed as a measure of the non-commutativity of the Hamiltonian at different time points. 
In practice, this is reflected by the increasing depth of nested commutators of $H(t)$ with increasing order of the perturbation series. 
Formally, the convergence of the expansion is assessed based on the expansion parameter 
$\lVert H\rVert/\omega$ with a submultiplicative matrix norm. 
We define 
\begin{subequations}
    \begin{align}
        \Lambda(t) &=: \sum_{m=1}^{\infty} \Lambda^{(m)}(t)\,,\quad \Lambda^{(m)}(t) \propto 1/\omega^m\,,  \\ 
        F &=: \sum_{m=1}^{\infty} F^{(m)} \,,\quad F^{(m)} \propto 1/\omega^{m-1} \,, 
    \end{align}
\end{subequations}
and their truncated counterparts as 
\begin{subequations}
    \begin{align}
        \Lambda^{[n]}(t) &:= \sum_{m=1}^{n} \Lambda^{(m)}(t)\,,  \\ 
        F^{[n]} &:= \sum_{m=1}^{n} F^{(m)}\,. 
    \end{align}
\end{subequations}
In the full propagator, $F$ is multiplied with the time difference $(t-t_0)$
assumed to be of the magnitude of the cycle time $T$, i.e., the inverse modulation frequency $\omega$. 
This is why $F^{(m)}$ and $\Lambda^{(m)}(t)$ scale differently in $1/\omega$. 

Usually \cite{casasFloquetTheoryExponential2001,blanesMagnusExpansionIts2009,manangaIntroductionFloquetMagnusExpansion2011}, 
Eq.~(\ref{eqn:FME_dgl}) is integrated by iteration in increasing order, 
which provides $\Lambda^{(m)}(t)$ in dependence of $F^{(m)}$, 
and $F^{(m)}$ is found by making use of $\Lambda^{(m)}(t_0+T)=\Lambda^{(m)}(t_0)$. 
For the first two orders, this yields 
\begin{subequations}
    \begin{align}
    \Lambda^{(1)}(t) &= \Lambda^{(1)}(t_0) + \frac{B_0}{0!} \int_{t_0}^{t} \text{d}t'\, \bigl(H(t') - F^{(1)}\bigr) \,, \\ 
    \Lambda^{(2)}(t) &= \Lambda^{(2)}(t_0) - \frac{B_0}{0!}\int_{t_0}^{t} \text{d}t'\, F^{(2)} 
            + i\cdot \frac{B_1}{1!} \int_{t_0}^{t} \text{d}t'\, \bigl[\Lambda^{(1)}(t'), H(t') + F^{(1)} \bigr] \,. 
    \end{align}
\end{subequations}

In this representation, the precise formulas for $F$ and $\Lambda$ depend on the choice of 
$\Lambda^{(m)}(t_0)$. 
This choice is not unique and can be considered a gauge freedom of FME. 
In particular, this means that the same temporal evolution operator $U(t,t_0)$ is produced
independently of the choice for $\Lambda(t_0)$. 
This is exactly true for the infinite series expansions, 
but not for their truncated variants, which only  yield perturbative agreement. 
As shown in the following subchapters, 
AHT is obtained by choosing $\Lambda(t_0)=0$, 
whereas FVV results from 
$\int_{0}^{T}\Lambda(t)dt=0$ being equivalent to 
setting the zeroth Fourier coefficient of the periodic kick operator~$\Lambda$ to zero (see Eq.~(\ref{eqn:LambdaFourier})). 
Consequently, the complete effective Hamiltonian and the kick operator of AHT and FVV are related by 
\begin{subequations}
    \begin{align}
        F_\text{AHT} &= e^{-i\Lambda_\text{FVV}(t_0)} F_\text{FVV} e^{+i\Lambda_\text{FVV}(t_0)}\,, \\
        \Lambda_\text{AHT}(t) &= i \log\Bigl[ e^{-i\Lambda_\text{FVV}(t)} e^{+i\Lambda_\text{FVV}(t_0)} \Bigr] \,, 
        \label{eqn:lambda_aht_fvv}
    \end{align}
    \label{eqn:AHT_FVV_relation}
\end{subequations}
leading to the same algebraic evolution operator $U_\text{AHT}(t,t_0)=U_\text{FVV}(t,t_0)$. 

As soon as the series expansion is truncated, AHT and FVV yield different results.
The relations of Eq.~(\ref{eqn:AHT_FVV_relation}) are not exactly true anymore 
but differ in the indicated order of the expansion,
i.e., 
\begin{subequations}
    \begin{align}
        F_\text{AHT}^{[n]} &= e^{-i\Lambda_\text{FVV}^{[n-1]}(t_0)} F_\text{FVV}^{[n]} e^{+i\Lambda_\text{FVV}^{[n-1]}(t_0)} + \mathcal{O}(1/\omega^{n}) \,, \\ 
        \Lambda_\text{AHT}^{[n]}(t) &= i \log\Bigl[ e^{-i\Lambda_\text{FVV}^{[n]}(t)} e^{+i\Lambda_\text{FVV}^{[n]}(t_0)} \Bigr] + \mathcal{O}(1/\omega^{n+1})
        \,. 
    \end{align}
\end{subequations}

\subsection{Average Hamiltonian Theory}

The following section outlines the fundamentals of the original derivation of AHT. 
The Magnus expansion solves the time-dependent Schrödinger equation for a general time-dependent, not necessarily periodic Hamiltonian $H(t)$, 
while preserving the unitary structure of the propagator at any stage of truncation. 
This is implemented using an exponential ansatz $U(t,t_0) = \exp\bigl[\Omega_{t_0}(t)\bigr]$
with an anti-Hermitian operator $\Omega_{t_0}(t)$. 
The exponent is expressed as a perturbative series that is truncated in the desired order, 
$\Omega_{t_0}(t)=\sum_{n=1}^{\infty}\Omega_{t_0}^{(n)}(t)$. 
This ansatz is inserted into the Schrödinger equation and 
leads to a differential equation for $\Omega_{t_0}(t)$, known as the Magnus theorem, 
\begin{equation}
    \frac{d\Omega_{t_0}}{dt} 
    = -i\sum_{k=0}^\infty \frac{B_k}{k!} [\Omega_{t_0}, H]_k 
    \stackrel{(\ref{eqn:BernoulliGenerating})}{=} -i\cdot f({\cal L}_{\Omega_{t_0}}) H \,. 
    \label{eqn:magnus_theorem}
\end{equation}
The signed Bernoulli numbers are, again, denoted by $B_k$. 
Next, the Hamiltonian is replaced via $H\rightarrow \lambda\cdot H$ 
with some small auxiliary parameter $\lambda$, which is set to $\lambda=1$ later. 
This allows one to specify $n$-th order terms in the expansion by 
$\Omega_{t_0}^{(n)}(t) \propto \lambda^n$. 
Based on that, Eq.~(\ref{eqn:magnus_theorem}) is integrated  order by order by iteration
together with the initial condition $\Omega_{t_0}^{(n)}(t_0)=0$. 
The first three differential equations and their solutions are 
\begin{subequations}\label{eqn:diff_eq_magnus}
    \begin{align}
        \Omega_{t_0}^{(1)}(t) &= -i \int_{t_0}^{t}\text{d}t' \, H(t')\,, \\ 
        \Omega_{t_0}^{(2)}(t) &= -i \int_{t_0}^{t} \text{d}t' \, \biggl( -\frac{1}{2} [\Omega_{t_0}^{(1)}(t'), H(t')] \biggr) \\ 
            &= -\frac{1}{2} \int_{t_0}^{t}dt_1 \, \int_{t_0}^{t_1}dt_2 \, [H(t_1), H(t_2)]\,, \\ 
        \Omega_{t_0}^{(3)}(t) &= -i \int_{t_0}^{t} \text{d}t' \, \biggl( -\frac{1}{2} [\Omega_{t_0}^{(2)}(t'), H(t')] + \frac{1}{12} \bigl[\Omega_{t_0}^{(1)}(t'),[\Omega_{t_0}^{(1)}(t'), H(t')]\bigr] \biggr)\\ 
            &= \frac{i}{6} \int_{t_0}^{t}dt_1 \,\int_{t_0}^{t_1}dt_2 \,\int_{t_0}^{t_2}dt_3 
            \Bigl\{ \bigl[H(t_1), [H(t_2), H(t_3)]\bigr] + \bigl[H(t_3), [H(t_2), H(t_1)]\bigr] \Bigr\} \,. 
    \end{align}
\end{subequations}
Consequently, $\Omega_{t_0}^{(n)}(t)$ leads to products consisting of $n$ Hamiltonians, i.e., 
$\Omega_{t_0}^{(n)}(t)\propto |H|^n (t-t_0)^n$. 
Formally, convergence is guaranteed for sufficiently small $\lVert H\rVert \cdot \lvert t-t_0\rvert$.  

Average Hamiltonian Theory (AHT), also known as coherent averaging  \cite{haeberlenCoherentAveragingEffects1968,haeberlenHighResolutionNMR2012}, 
applies the Magnus expansion to time-periodic Hamiltonians
and focuses on the stroboscopic evolution over integer multiples of the period $T$. 
For periodic systems, the Floquet theorem~(\ref{eqn:floquettheorem}) 
implies the convenient asset that 
\begin{equation}
    U(t_0+nT,t_0)=\bigl(U(t_0+T,t_0)\bigr)^n\,, \, n\in \mathbb{Z}\,.
\end{equation}
Thus, it makes sense to calculate only the single-cycle evolution operator $U(t_0+T,t_0)$  
with the help of the Magnus expansion 
and use it to determine the state of the 
system at stroboscopic time points $t=t_0+nT$, $n\in \mathbb{Z}$. 
This forms a solid base for investigating the dynamics in the course of time 
because $T$ is in the usual applications a small parameter. 

Since the Hamiltonian is periodic in time, the results of the Magnus expansion 
can be made explicit by inserting the Fourier 
series $H(t) = \sum_n H_n e^{in\omega t}$ in Eq.~(\ref{eqn:diff_eq_magnus}), 
see, for instance, Ref.\ \cite{leskesFloquetTheorySolidstate2010a}. 
This allows one to solve the integrals for a general periodic $H(t)$, 
but requires the determination of coefficients in $n$-th order of the form 
\begin{equation}
    C_{\{m_i\}} := \int_{t_0}^{t_0+T}dt_1 \int_{t_0}^{t_1}dt_2 \dots \int_{t_0}^{t_{n-1}}dt_n e^{i\omega (m_1t_1 + m_2t_2 + \dots + m_nt_n)} 
\end{equation}
with $\{m_i\}$ being integers. 
Although this procedure is accurate, 
it is complicated and prone to errors. 
A growing number  
of special cases for the integers $m_i$, $i=1,2,\dots,n$, needs to be distinguished, 
making the calculation by pen and paper in third order non-trivial 
and in fourth order almost impossible. 
In our opinion, it is easier to follow the procedure presented in App.~\ref{app:calc_scheme},  
where we switch to the Fourier basis, avoiding explicit integration. 

The final single-cycle evolution is described by $\exp\bigl[\Omega_{t_0}(t_0+T)\bigr]$. 
Taking Eq.~(\ref{eqn:stroboscopicHamiltonian}) into account, 
we easily see that FME is related to AHT in each order $n$ by 
\begin{equation}
    \tilde{F}^{(n)} = F^{(n)} = \frac{i}{T} \Omega_{t_0}^{(n)}(t_0+T)\,\,, \,\Lambda^{(n)}(t_0) = 0\,. 
\end{equation}

Some computational routines are more easily implemented using the integral formulas directly, 
which are included above for the first three orders and in App.~\ref{app:4thorder} in  fourth order for completeness.
The explicit solution based on the Fourier coefficients of the time-dependent Hamiltonian is most convenient for Hamiltonians with a limited number of Fourier coefficients, e.g., the Hamiltonian generated by sample rotation that contains only Fourier coefficients from $n=-2$ to $n=2$. For time-dependent Hamiltonians with sudden changes in phase or amplitude as they are typically generated by pulse sequences, the integral formulas are often easier to evaluate since an infinite number of Fourier coefficients would have to be considered while the number of time periods with constant phase and amplitude are finite.

\subsection{Floquet--Van Vleck approach}

\begin{figure*}
    \includegraphics[width=\textwidth]{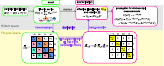}
    \caption{\label{fig:FloquetTheory}
        Overview of the Floquet--Van Vleck approach. 
        The transformation of the Schrödinger equation allows one to rewrite the differential equation as an eigenvalue-eigenvector problem of an infinite-dimensional matrix, the Floquet Hamiltonian $\bar{H}_\mathcal{F}$. 
        It acts in the Floquet space, which is spanned by the direct product of the Hilbert space basis and the Fourier states $\{\lvert n\rangle\}$. 
        The Fourier indices of the entries in the central block $\langle n\lvert \dots\rvert n\rangle$ sum up to zero, i.e., they are secular terms. 
        Non-secular terms, of which the Fourier indices sum  to $m\neq 0$, are on the non-block-diagonals $\langle n\lvert \dots\rvert n-m\rangle$. 
        The Van Vleck perturbation theory shifts the effect of the fast sidebands to the central band. 
        In order to compute the full stroboscopic dynamics, the effective Hamiltonian $\mathcal{H}_\text{eff}$
        has to be back-rotated in order to operate in the correct basis. 
        This provides the  stroboscopic Hamiltonian $\mathcal{H}_\text{s}$. 
    }
\end{figure*}

In contrast to AHT, Floquet theory \cite{floquetEquationsDifferentiellesLineaires1883}
is exclusively tailored to time-periodic Hamiltonians. 
For the full derivation, 
we refer the reader to existing literature \cite{shirleySolutionSchrodingerEquation1965,leskesFloquetTheorySolidstate2010a,scholzOperatorbasedFloquetTheory2010a,IVANOV202117}. 
In the following, we provide a summary combined with 
the illustrated overview in Fig.~\ref{fig:FloquetTheory}. 

The essential idea is to perform a Fourier expansion of the Schrödinger equation  
and make use of the implemented Fourier basis, which is associated to the corresponding Fourier sidebands. 
Formally, this defines a homomorphism 
mapping the time-dependent differential equation in Hilbert space 
to an eigenvalue-eigenvector problem of an infinite-dimensional matrix, 
which is known as the Floquet Hamiltonian.
It corresponds to the Floquet space, which is formed by the direct product of the Hilbert space basis 
and the infinite-dimensional Fourier basis. 
With the help of the Van Vleck perturbation theory, 
the Floquet Hamiltonian is block-diagonalized up to a desired order in $1/\omega$. 
The effect of the fast sidebands at $m\omega$ ($m\neq 0$) is folded back to the block-diagonal band at $m=0$. 
Consequently, the diagonal blocks of the transformed Floquet Hamiltonian yield 
the desired effective Hamiltonian $\mathcal{H}_\text{eff}$. 
The unitary Van Vleck transformation in Floquet space corresponds to a unitary transformation of $\mathcal{H}_\text{eff}$ 
in Hilbert space as well. 
This basis change must be taken into account if one aims at the time evolution of the original problem. 
An additional rotation $R$ in Hilbert space yields the correctly tilted Hamiltonian operator, 
typically referred to as the stroboscopic Hamiltonian $\mathcal{H}_\text{s}=R \mathcal{H}_\text{eff} R^\dagger$. 
It governs the stroboscopic dynamics, i.e., 
\begin{equation}
    U(t_0+T,t_0) = e^{-i\mathcal{H}_\text{s}\cdot T} = R e^{-i\mathcal{H}_\text{eff}\cdot T} R^\dagger\,. 
\end{equation}

In the context of FVV, terms such as  
\begin{equation}
    \mathbf{H}_{m,\{n_i\}} :=
    {H_{n_1}H_{n_2}\dots H_{n_k}H_{m-n_1-n_2-\dots-n_k}} 
    \label{eqn:secular}
\end{equation}
are called 
secular if $m=0$ and non-secular if $m\neq 0$. 
FVV is also called secular averaging because the diagonal blocks of the Floquet Hamiltonian 
contain exclusively secular terms. 
Consequently, the same holds true for the effective Hamiltonians $\mathcal{H}_\text{eff}$, 
and the fast, recurring micromotion exclusively appears in the rotation operator $R$ in the form 
of non-secular terms. 
Only if $\mathcal{H}_\text{eff}$ and $R$ are combined yielding $\mathcal{H}_\text{s}$, 
secular terms are mixed with non-secular ones. 
At this point, we emphasize that the eigenvalues of $U(t_0+T,t_0)$ are completely determined by the effective 
Hamiltonian $\mathcal{H}_\text{eff}$ and are not changed by the unitary transformation $R$. 
The latter only changes the intensity of the transitions and is only needed if one is interested in the full stroboscopic evolution. 

If the single-cycle evolution operators $U(t_0+T,t_0)$ of FVV and FME are compared, the relations 
\begin{equation}
    \mathcal{H}_\text{eff} = F\,,\quad 
    \mathcal{H}_\text{s} = \tilde{F}\,,\quad 
    R = e^{-i\Lambda(t_0)}\,, 
\end{equation}
hold. 
A characteristic of $F$ in FME is that the sum $m$ of the indices within a product, as in Eq.~(\ref{eqn:secular}), 
is always connected to 
a factor $e^{im\omega t_0}$. 
Consequently, one realizes two crucial properties of FVV: 
First, because $\mathcal{H}_\text{eff}$ contains exclusively secular (i.e., $m=0$) terms, 
FVV is the only version of FME with an $F$ that is independent of $t_0$. 
We stress that this remains true at any stage of truncation. 
As becomes even clearer with the background given in App.~\ref{app:calc_scheme}, 
the same applies to the kick operator, i.e., $\partial_{t_0}\Lambda^{(n)}(t)=0\,\forall n\in \mathbb{N}$.  
Second, all non-secular (i.e., $m\neq 0$) terms are collected in $\Lambda$ and 
linked to the $m$-th sideband of the periodic $\Lambda(t)$. 
Consequently, $\Lambda$ exclusively consists of non-secular terms. 
Hence, FVV is obtained within FME if 
the zeroth Fourier coefficient of $\Lambda$ is set to zero, 
\begin{equation}
    \int_{0}^{T}dt\,\Lambda(t) = 0 \,, 
\end{equation}
which implicitly fixes the initial condition $\Lambda(t_0)$ and thereby $\Lambda(t)$ and $F$. 

Although FVV might appear less amenable than AHT, there are two main advantages in its use. 
First, the location of resonances in NMR spectra depends only on the eigenvalues of $F$, 
independent of $t_0$. 
This is guaranteed by FVV at any truncation stage due to the fact that $F$ is independent of $t_0$. 
Second, FVV is advantageous on the practical level because
there are far fewer terms. This will be discussed in detail below.

\section{Overview of relevant orders}
\label{sec:orders}

The following serves as a reference for $F$ and $\Lambda_m$ 
in the commutator form for FVV and AHT, respectively. 
The commutator form is useful for all time-periodic Hamiltonians $H(t)=H(t+T)$, for which the Fourier series 
is defined in Eq.~(\ref{eqn:fourier_H}). 
Our results are based on the integration-free method, which is discussed in detail in 
App.~\ref{app:calc_scheme}.
There is an error in the third-order AHT expression $F^{(3)}$ in \cite{leskesFloquetTheorySolidstate2010a}. 
We are also able to present the fourth order of both approaches, which is usually considered to be too cumbersome to be derived.

As a benchmark, we numerically validate both eigenvalues and eigenvectors of the stroboscopic propagator 
$U(t_0+T,t_0)$ for random and simple non-random time-periodic Hamiltonians $H(t)$ in several dimensions. 
We show that the deviations of both theories to the numerically exact propagator scale as expected 
from their truncation level. This validates the analytical results numerically. 
The outcomes as well as the technical details are briefly presented in the conclusion of this section.
In addition, we also compared the formulas to the analytical integration of the time-dependent Hamiltonian for the first four orders using Mathematica calculations for a dipolar-coupled two- and three-spin system under MAS, shown in the SI Sect.~S5.  

Notice that the $n$-th order evolution operator $U^{[n]}(t_0+T,t_0)$ corresponds to 
the $n$-th order of $\tilde{F}^{[n]}=e^{-i\Lambda^{[n-1]}(t_0)} F^{[n]} e^{+i\Lambda^{[n-1]}(t_0)} $. 
In AHT, $\tilde{F}^{[n]}=F^{[n]}$ holds true in contrast to FVV where
the first $n$ orders of $F$ and the $n-1$ orders of $\Lambda$ are needed
to capture all terms in $n$-th order of $\tilde{F}$. 
The kick operator does not appear on its own, only in products.  
This is evident when examining the second order as an example 
with the help of the Hadamard lemma \cite{wilcoxExponentialOperatorsParameter1967} implying
\begin{subequations}
    \begin{align}
        \tilde{F}^{[2]} 
        &\stackrel{\text{Had.}}{=} 
        \sum_{n=0}^{\infty} \frac{1}{n!} [-i\Lambda^{[1]}(t_0), F^{[2]}]_n 
        \\ 
        &= F^{[2]} -i [\Lambda^{(1)}(t_0), F^{(1)}] + \mathcal{O}(1/\omega^2) 
        \,. 
    \end{align}
\end{subequations}
Based on this consideration, this chapter includes the first three orders of $F$ and two orders of $\Lambda_m$, being the $m$-th Fourier coefficient of the periodic kick operator $\Lambda(t)$. 
The subsequent fourth order of $F$ and third order of $\Lambda_m$ are given in App.~\ref{app:4thorder}. 

For clarity, we define the short notation 
\begin{equation}
    [n,m] := [H_n,H_m]\,. 
\end{equation}

We want to briefly sketch the integration-free method with details in 
App.~\ref{app:calc_scheme}. 
The essential idea is to exploit the periodicity of the kick operator $\Lambda(t)$ by applying a Fourier transformation 
to the defining differential Eq.~(\ref{eqn:FME_dgl}). 
Due to the orthogonality of the Fourier basis, the $e^{im\omega t}$-prefactors need to be the same, 
which leads to 
\begin{equation} 
    \begin{aligned}
        im\omega\cdot \Lambda_m
            &= -F \cdot\delta_{m,0} + H_m \\
            &+ \frac{i}{2} \Bigl\{ \sum_{n_1} [\Lambda_{n_1}, H_{m-n_1}] 
            + [\Lambda_{m}, F]  \Bigr\} \\ 
            &+ \sum_{j=2}^\infty (-i)^j \frac{B_j}{j!} \biggl\{ \sum_{n_1,\dots,n_j} \Bigl[\Lambda_{n_1},\bigl[\dots,[\Lambda_{n_j}, H_{m-n_1-\dots-n_j}]\dots \bigr]\Bigr] \\ 
            &+ (-1)^{j+1} \sum_{n_1,\dots,n_{j-1}} \Bigl[\Lambda_{n_1},\bigl[\dots,[\Lambda_{m-n_1-\dots-n_{j-1}} , F]\dots\bigr]\Bigr]  \biggr\} \,. 
    \end{aligned}
\end{equation}
In the next step, terms of the same order are iteratively equated, i.e., 
\begin{subequations}
    \begin{align}
        F^{(1)} &= H_0 \,, \\ 
        \Lambda_{m\neq 0}^{(1)} &= \frac{H_m}{im\omega} \,, \\ 
        F^{(2)} &= \frac{i}{2} \Bigl\{ \sum_{n_1} [\Lambda_{n_1}^{(1)}, H_{-n_1}]
        + [\Lambda_{0}^{(1)}, F^{(1)}] \Bigr\} \\ 
        &= \sum_{n\neq 0} \frac{[n, {-n}]}{2n\omega}
        + i\cdot [\Lambda_{0}^{(1)}, H_0] \,, \\ 
        \Lambda_{m\neq 0}^{(2)} &= \frac{1}{2m\omega} \Bigl\{
            \sum_{n_1} [\Lambda_{n_1}^{(1)}, H_{m-n_1}] 
            + [\Lambda_{m}^{(1)}, F^{(1)}]  
            \Bigr\} \\ 
            &= 
            \sum_{n\neq 0} \frac{[n, {m-n}] }{2inm\omega^2}
            + \frac{[m, 0]}{2im^2\omega^2} 
            + \frac{[\Lambda_{0}^{(1)}, H_{m}]}{2m\omega} \,. 
    \end{align}
\end{subequations}
We realize that the choice of $\Lambda(t_0)$ actually translates into a choice of the zeroth Fourier coefficient $\Lambda_0$, which here is an operator.

\subsection{Floquet--Van Vleck theory}

In this gauge choice meaning $\Lambda_0^{(n)}\equiv 0$ in each order $n$,
the effective Hamiltonian $F$ contains all secular terms, i.e., the sum of the indices of the Fourier coefficients of $H(t)$ in a product vanishes. 
All non-secular terms arise in the kick operator $\Lambda$. 

\begin{subequations}
    \begin{align}
        F^{(1)} &= H_0 \\ 
        F^{(2)} &= \sum_{n\neq 0} \frac{[n,-n]}{2n\omega} \\ 
        F^{(3)} &= \sum_{n\neq 0} \sum_{m\neq 0} \frac{1}{3nm\omega^2} \bigl[n,[m,-m-n]\bigr] 
            + \sum_{n\neq 0} \frac{1}{6n^2\omega^2} \bigl[n,[0,-n]\bigr] 
    \end{align}
\end{subequations}

\begin{subequations} 
    \begin{align} 
        \Lambda^{(N)}(t) &= \sum_{m\neq 0} \Lambda_m^{(N)} e^{im\omega t} \\ 
        \Lambda_{m\neq 0}^{(1)} &= \frac{H_m}{im\omega} \\ 
        \Lambda_{m\neq 0}^{(2)} &= \sum_{n\neq 0} \frac{[n,m-n]}{2inm\omega^2} 
        + \frac{[m,0]}{2im^2\omega^2} 
    \end{align}
\end{subequations}

\subsection{Average Hamiltonian theory}

In principle, the effective Hamiltonian $F$ of AHT is exactly the same as the stroboscopic Hamiltonian $\tilde{F}$ in FVV. 
However, for their truncated expansions, the equality reduces to $F^{[n]}_\text{AHT}=\tilde{F}^{[n]}_\text{FVV} + \mathcal{O}(1/\omega^{n})$. 
In the following presentation, we do not implicitly set $t_0=0$ since it makes one of the main differences between FVV and AHT visible. 
The $t_0$-dependence is directly connected to non-secular terms which we deem to be unfortunate components of the effective Hamltonian. 
The gauge of AHT, $\Lambda^{(n)}(t_0)\equiv 0$, translates into $\Lambda_0^{(n)}\equiv -\sum_{m\neq 0}\Lambda_m^{(n)}e^{im\omega t_0}$ 
for the zeroth Fourier coefficient in each order $n$. 

\begin{subequations}
    \begin{align}
        F^{(1)} &= H_0 \\
        F^{(2)} &= \sum_{n\neq 0} \frac{[n,-n]}{2n\omega} + \sum_{n\neq 0}\frac{[0,n]}{n\omega }e^{in\omega t_0} \\ 
        F^{(3)} &= \sum_{n\neq 0}\sum_{m\neq 0} 
            \frac{ \bigl[n,[m,-m-n]\bigr] }{3nm\omega^2}  
            + \sum_{n\neq 0}\sum_{m\neq 0} 
            \frac{ \bigl[m,[-n,n]\bigr] }{2nm\omega^2} e^{im\omega t_0} 
            \\  
            &+ \sum_{m\neq 0} \sum_{n\neq -m} 
            \frac{ \bigl[n,[m,0]\bigr] }{m(m+n)\omega^2} e^{i(n+m)\omega t_0}   
            + \sum_{n\neq 0} 2 \frac{ \bigl[n,[0,-n]\bigr] }{3n^2\omega^2}  
    \end{align}
\end{subequations}

\begin{subequations}
    \begin{align} 
        \Lambda^{(N)}(t) &= \sum_{m\neq 0}\Lambda_m^{(N)} (e^{im\omega t} - e^{im\omega t_0}) \\ 
        \Lambda_{m\neq 0}^{(1)} &= \frac{H_m}{im\omega} \\ 
        \Lambda_{m\neq 0}^{(2)} &= \sum_{n\neq 0} \frac{[n,m-n]}{2inm\omega^2} 
        + \sum_{n\neq 0} \frac{[m,n]}{2inm\omega^2} e^{in\omega t_0} 
        + \frac{[m,0]}{2im^2\omega^2} 
    \end{align}
\end{subequations}

\subsection{Numerical validation}
\label{sub:numericalvalidation}

The main objective of the following section is to validate the formulas numerically to avoid oversights in view of their complexity. 
For this, we define time-periodic Hamiltonians for several systems and calculate 
their $n$-th order single-cycle evolution operators $U^{[n]}(t_0+T,t_0)$ in three ways: 
\begin{enumerate}
    \item[1)] AHT $e^{-iF^{[n]}_\text{AHT}\cdot T}$, 
    \item[2)] complete FVV, i.e., with kick operator $e^{-i\Lambda^{[n-1]}_\text{FVV}(t_0)}e^{-iF^{[n]}_\text{FVV}\cdot T}e^{+i\Lambda^{[n-1]}_\text{FVV}(t_0)}$, 
    \item[3)] FVV without kick operator $e^{-iF^{[n]}_\text{FVV}\cdot T}$. 
\end{enumerate}
We add the last case 
because it is common practice in NMR applications to take 
into account just the effective Hamiltonian $\mathcal{H}_\text{eff}=F$  
without the additional rotation $R=\exp(-i\Lambda(t_0))$. 

\begin{figure*}
    \includegraphics[width=0.95\textwidth]{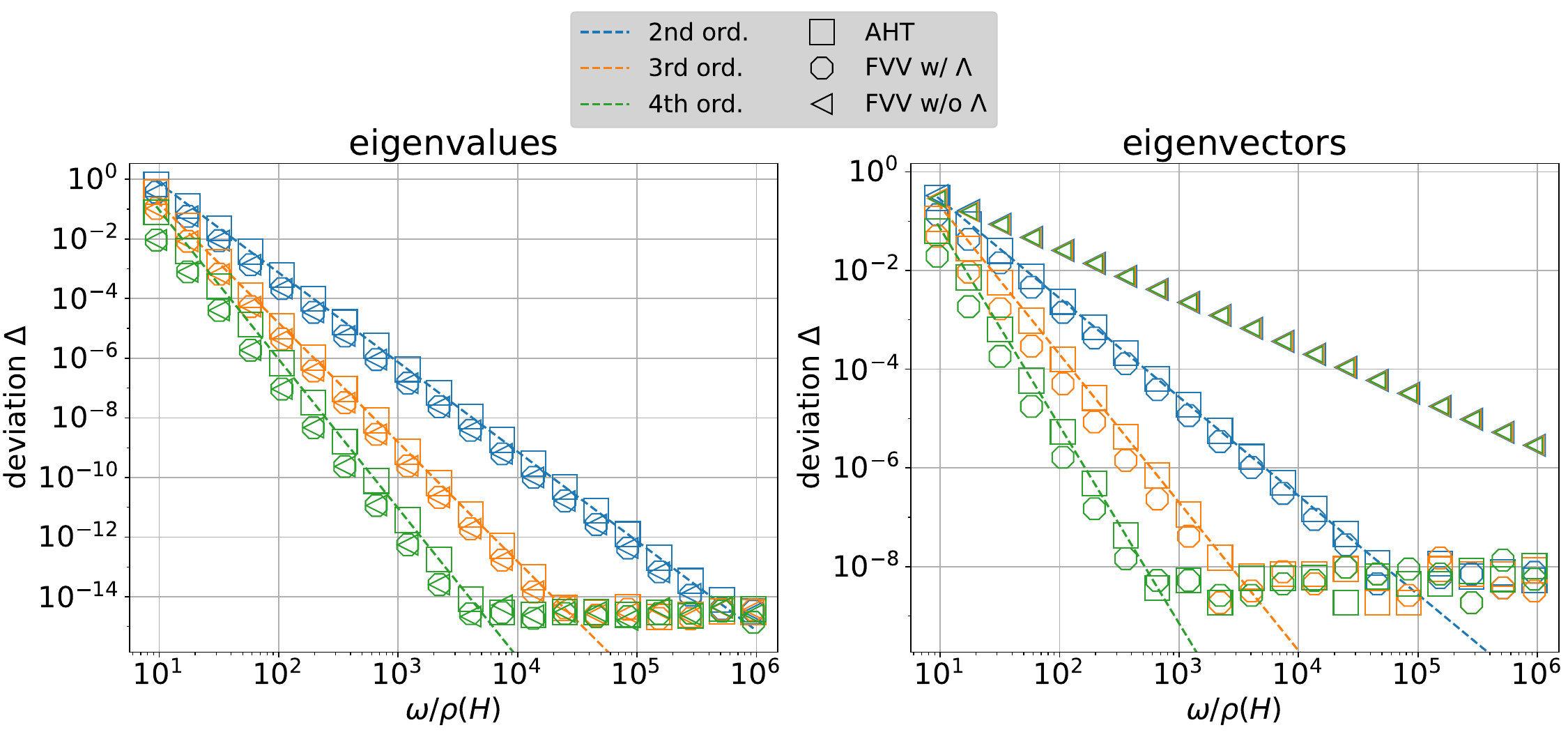} 
    \caption{
        The deviations of $U(t_0+T,t_0)$ in AHT, FVV with and without the kick operator $\Lambda$ (often referred to the stroboscopic and effective Hamiltonian in Floquet theory, respectively) 
        to the numerically exact time propagator $U(t_0+T,t_0)$ 
        are illustrated for a random $8\times 8$ Hamiltonian. 
        The dashed lines indicate the expected scaling behavior according to the truncated order in the perturbation parameter. 
        In $n$-th order, this is $(\rho(H)/\omega)^{n}$ for the eigenframe and $(\rho(H)/\omega)^{n+1}$ for the eigenvalues. 
        Both AHT and FVV scale as expected, however, AHT is about three times worse than FVV
        as  can be seen in Fig.~\ref{fig:numerical_validation_factor}.
        Clearly, the kick operator is needed in FVV in order to operate in the correctly tilted basis. 
    }
    \label{fig:numerical_validation}
\end{figure*}

The error of the three different implementations with respect to the eigenvalues $\{\kappa_{i,j}\}_j$ and 
the eigenvectors $\{\vec{v_{i,j}}\}_j$ of $U^{[n]}(t_0+T,t_0)$ is measured by 
\begin{subequations}
    \begin{align}
        \Delta^\text{val}_i & := \sum_{j=1}^{d} {\lvert \kappa_{i,j}-\kappa_{\text{true},j}\rvert} \,, \\ 
        \Delta^\text{vec}_i &:= \sum_{j=1}^{d}  \arccos(\lvert \langle \vec{v}_{\text{true},j} \lvert \vec{v}_{i,j}\rangle \rvert ) \,, 
    \end{align}
\end{subequations}
making use of the numerically exact evolution operator. We investigate 
several systems ranging from simple spin systems to random Hermitian operators 
in 2, 4, 8 and 16 dimensions.  
The deviations of all systems scale correctly according to the approximated order,
as shown in Fig.~\ref{fig:numerical_validation}. 
We normalize the frequency $\omega$ by the spectral radius $\rho(H)\leq\lVert H\rVert$ which is a lower bound for the sub-multiplicative norm of a quadratic matrix. 
Technical details on both the error functions and the specific systems are found in App.~\ref{app:toysystems}. 

In addition to the correctness of our formulas, 
the results clearly show that only the propagator with kick operator yields the correct eigenframe. 
We conclude that the kick operator is crucial for mathematical rigor, except for the eigenvalues.  

\begin{figure*}
    \includegraphics[width=0.95\textwidth]{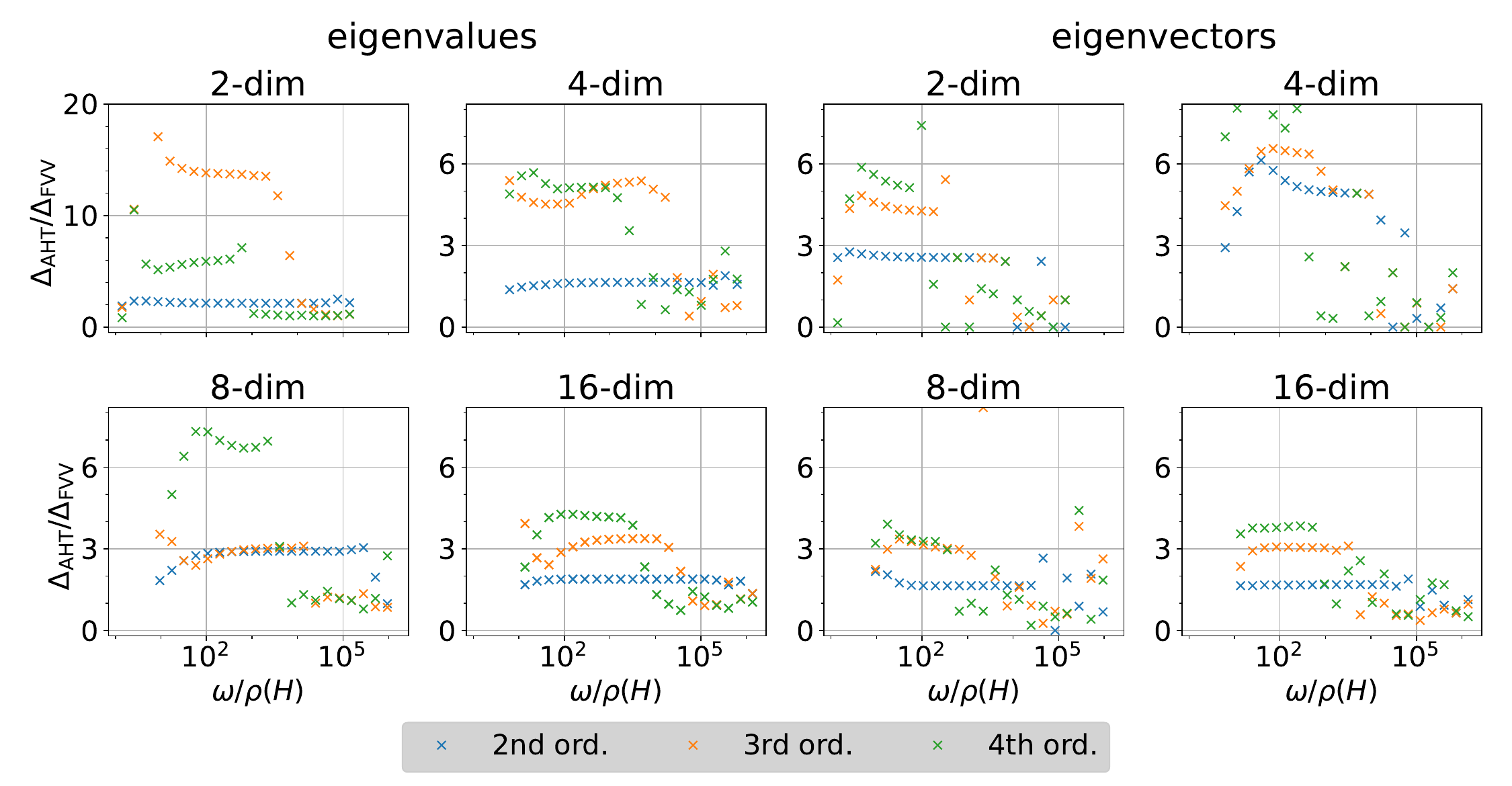} 
    \caption{
        Both AHT and FVV yield the expected scaling behavior in the perturbation parameter. 
        However, FVV performs better than AHT in the approximation of the stroboscopic evolution operator $U(t_0+T,t_0)$. 
        AHT deviates $\sim 3.7$ times more than FVV in the eigenvalues, and $\sim 2.6$ in the eigenvectors. 
    }
    \label{fig:numerical_validation_factor}
\end{figure*}

Furthermore, we note that the deviation is expected to follow $\Delta = c \omega^{-a}$, with some constant $c$, and $a$ is predicted by the truncation level. 
It is remarkable that, for all considered systems, FVV provides slightly more accurate approximations as AHT 
because $c$ is consistently smaller for FVV. 
This can be seen in Fig.~\ref{fig:numerical_validation_factor}, in which we show the ratio of the deviations between 
AHT and FVV, 
which is on average $\Delta_\text{AHT}/\Delta_\text{FVV}\approx 3.1$. 

\section{NMR applications}
\label{sec:NMR_appli}

The following section serves both as an illustration for the theoretical framework 
presented here and for drawing important conclusions for NMR-related applications
and the simulation of spectra. 
Unless otherwise mentioned, 
the latter is based on stroboscopic observation with the help of the single-cycle evolution operator $U(t_0+T,t_0)$, 
which is computed using four different approaches: 
\begin{enumerate}
    \item[a)] numerically exact integration,
    \item[b)] AHT, 
    \item[c)] complete FVV (F w/ $\Lambda$), 
    \item[d)] FVV without kick operator (F w/o $\Lambda$). 
\end{enumerate}
We include the last case because it is common practice in the NMR community to take only the effective 
Hamiltonian $F$ into account, neglecting the kick operator $\Lambda(t_0)$. 

The NMR spectra, i.e., the signal in the frequency domain $\mathcal{S}(\omega)$ with $\omega = 2\pi\nu$ corresponding 
to the time-dependent signal $\mathcal{S}(t)$ of the according detection operator $\mathcal{S}$ at stroboscopic time points, are calculated by 
\begin{equation}
    \mathcal{S}(\omega) = \sum_{jk} \rho_{jk} \mathcal{S}_{kj} \frac{1}{1 - e^{-\eta \cdot T} e^{i (\omega -\Delta \lambda_{jk})\cdot T}} \,. 
    \label{eqn:spectrum_summary}
\end{equation}
The sum is taken over the eigenstates of $U(t_0+T,t_0)$, which has the same eigenbasis as $\tilde{F}$. 
This means that the indices of $\mathcal{S}$ and of 
the initial state $\rho$ correspond to the matrix entries in the eigenbasis of the stroboscopic Hamiltonian. 
Resonances are located at $\omega = \Delta \lambda_{jk} :=\lambda_j-\lambda_k$ 
with $\{\lambda_j\}$ being the eigenvalues of the effective Hamiltonian $F$. 
An exponential line broadening $\eta$ is introduced in order to transform $\delta$-peaks into Lorentzian lines with a full width at half height of $2\eta$ in angular frequencies. 
Further details are provided in the SI Sect.~S1. 

We start with the Bloch--Siegert shift, 
a single spin $S=1/2$ system exposed to a strong magnetic field and a resonant radio-frequency field. 
This is followed by a homonuclear spin system with chemical shifts and dipolar coupling under
Magic-Angle Spinning (MAS). 
We consider two- and three-spin systems, 
both as single crystal and as powder sample, respectively. 
While these two use cases are based on stroboscopic sampling, the third example focuses on additional non-stroboscopic measurements in MAS 
because they more accurately reflect the actual laboratory setup and make the calculation of sidebands possible.

We recall that the Hamiltonians are given not in energy units, but in units of angular frequency. 
Spin operators are defined by $S^i = \sigma^i / 2$ with $\sigma^i$ being the
Pauli matrices \cite{pauliZurQuantenmechanikMagnetischen1927}. 
The corresponding raising and lowering operators are denoted by $S^\pm = S^x \pm iS^y$.

\subsection{Bloch--Siegert shift} 
\label{sub:blochsiegert}

A single spin-$1/2$ system with Zeeman interaction is considered. 
A resonant linear-polarized radio-frequency field enables transitions between the two levels. 
The Hamiltonian in the laboratory frame (LAB) reads 
\begin{equation}
    H_\text{LAB}(t) = \omega_0 S^z + 2 \omega_1 \cos(\omega_0 t) S^x \,, 
\end{equation}
in which $\omega_1/\omega_0\ll 1$ holds. 
As we intend to approximate the propagator in the limit $\lVert H\rVert/\omega_0 \ll 1$, 
we switch to the interaction frame 
$H_\text{I} = V H_\text{LAB} V^\dagger + i \dot{V}V^\dagger$
by means of the rotation $V(t) = e^{i\omega_0 S^z t}$ 
\begin{subequations}
\begin{align}
        H_\text{I}(t) &= \omega_1 S^x + \frac{\omega_1}{2} (S^+ e^{2i\omega_0t} + S^- e^{-2i\omega_0t}) \\
        &= \sum_{n=-2}^{2} H_n e^{in\omega_0t} \\ 
        H_0 &= \omega_1S^x\,,\quad 
        H_{\pm 1} = 0 \,,\quad 
        H_{\pm 2} = \frac{\omega_1}{2} S^{\pm} \,. 
\end{align}
\end{subequations}
Then, the perturbation parameter $\lVert H_\text{I}\rVert / \omega_0 \propto \omega_1/\omega_0$ is easy to control and indeed small. 
We assume an initial state of $\rho=\mathbb{I}/2+c\cdot S^z$ with a constant $c\in\mathbb{R}$, $\lvert c\rvert \leq 1$. 
The identity matrix cancels out in the trace of the quantum mechanical average $\langle\mathcal{S}(t)\rangle$, and the constant $c$ does not affect the spectrum because it is normalized. 
Hence, in the following, we just write $\rho=S^z$. 
Moreover, we measure along the $x$-axis, 
i.e., the detection operator is $\mathcal{S}=S^x$. 
 
For stroboscopic measurements at $t=t_0+nT$, $n\in\mathbb{N}^0$, in the $x$-direction, 
there is no difference between the LAB and the rotating frame. 
This can be seen from 
\begin{subequations}
    \begin{align}
        U_\text{LAB}(t_0+nT,t_0) &= V(nT) U_\text{I}(t_0+nT,t_0) \\ 
        V(nT) S^x V^\dagger(nT) &= S^x \\ 
        \Rightarrow \text{Tr}\{ U_\text{LAB}^{\phantom{\dagger}} \rho U_\text{LAB}^\dagger S^x \} 
        &= \text{Tr}\{ U^{\phantom{\dagger}}_\text{I} \rho U^\dagger_\text{I} S^x \} \,. 
    \end{align}
\end{subequations}

The first-order Hamiltonian $F^{(1)}_\text{AHT}=F^{(1)}_\text{FVV}=\omega_1S^x$ 
describes the precession of the spin around the radio-frequency field in the rotating frame. 
This corresponds to a complex spiral motion in the original LAB basis, 
and resonances are located around $\omega = 0,\,\pm\omega_1$ in first order according to the eigenvalues 
of the effective Hamiltonian. 
Higher-order contributions manifest in a slight shift
measured experimentally
and known as the Bloch--Siegert shift \cite{blochMagneticResonanceforNonrotatingField1940}. 

In second order, it is 
$F^{(2)}_\text{AHT}=-\frac{\omega_1^2}{4\omega_0}S^z$ and
$F^{(2)}_\text{FVV}=\frac{\omega_1^2}{4\omega_0}S^z$, and 
the kick operator in FVV is given by $\Lambda^{(1)}_\text{FVV}(0) = \frac{\omega_1}{2\omega_0} S^y$. 
Consequently, the full evolution operator in FVV is governed by 
the stroboscopic Hamiltonian 
\begin{subequations}
    \begin{align}
        \tilde{F}^{[2]}_\text{FVV} &= 
        e^{-i\Lambda^{[1]}_\text{FVV}(0)} F^{[2]}_\text{FVV} e^{i\Lambda^{[1]}_\text{FVV}(0)} \\
        &= \cos\Bigl(\frac{\omega_1}{2\omega_0}\Bigr) \Bigl\{ \omega_1 S^x + \frac{\omega_1^2}{4\omega_0} S^z \Bigr\} 
            + \sin\Bigl(\frac{\omega_1}{2\omega_0}\Bigr) \Bigl\{ -\omega_1 S^z + \frac{\omega_1^2}{4\omega_0} S^x \Bigr\} \\ 
        &= F^{[2]}_\text{AHT} + \omega_1\cdot\mathcal{O}\Bigl(\frac{\omega_1^2}{\omega_0^2}\Bigr) 
        \,, 
        \label{eqn:bloch_siegert}
    \end{align}
\end{subequations}
which is consistent with AHT in the approximated order. 
According to the eigenvalues of $F^{[2]}$ in both AHT and FVV, resonances are predicted to be at 
$\omega = 0,\,\pm\omega_1\sqrt{1+\frac{\omega_1^2}{16\omega_0^2}}$
yielding the anticipated additional shift. 
\begin{figure*}
    \centering
    \includegraphics[width=0.85\textwidth]{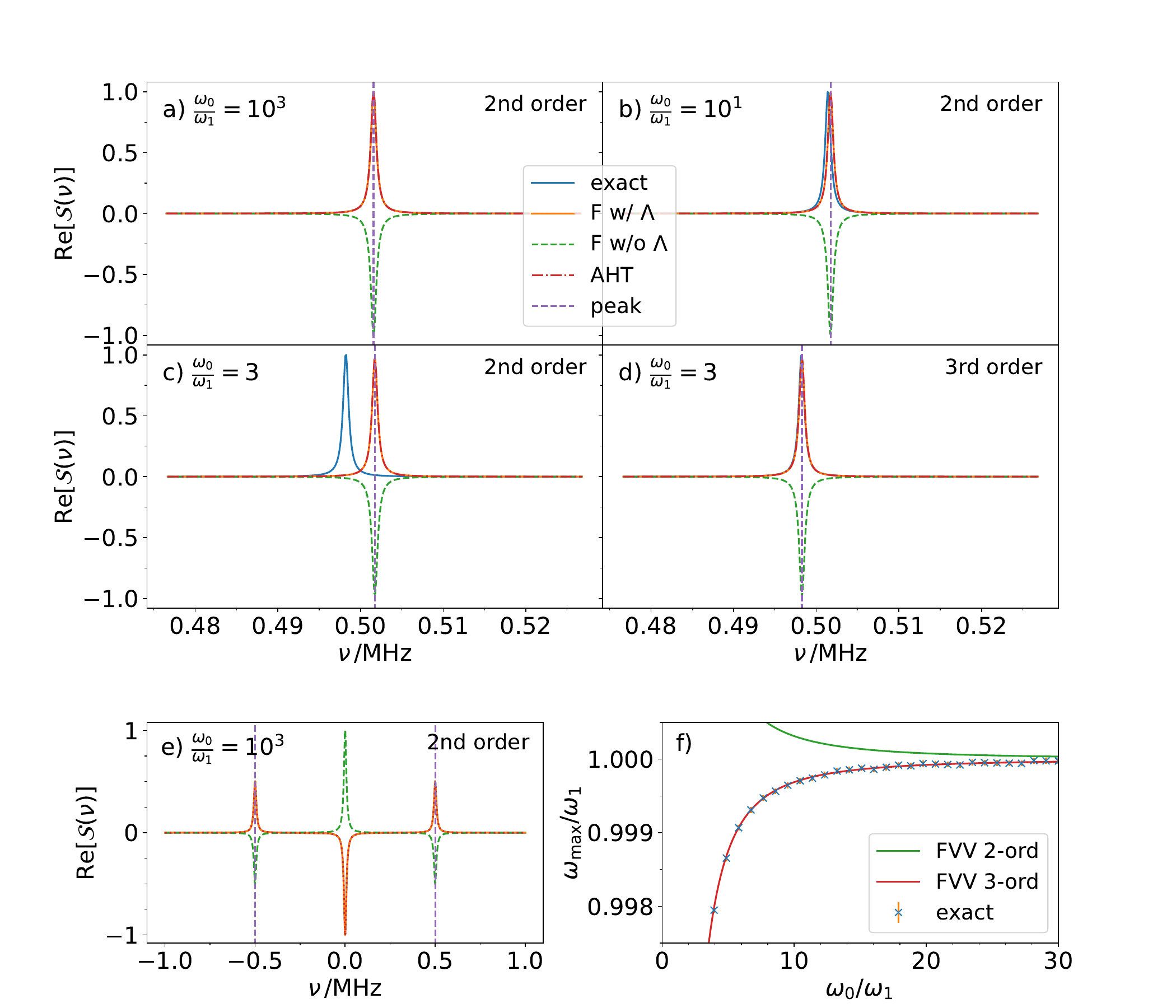}
    \caption{
        Bloch--Siegert shift \cite{blochMagneticResonanceforNonrotatingField1940} 
        of a single-spin system under a linear-polarized  resonant radio-frequency (rf) field. 
        Results according to AHT and FVV with and without kick operator $\Lambda$ are compared to the numerically exact outcome. 
        The detection operator is $\mathcal{S}=S^x$, and the initial density operator is $\rho=S^z$. 
        The amplitude of the rf field is set to $\omega_1/(2\pi)=0.5\,\mathrm{MHz}$, 
        and the resonance frequency $\omega_0$ is varied. 
        The broadening parameter is $\eta/(2\pi)=0.4\,\mathrm{kHz}$ in a)-d), and $\eta/(2\pi)=8\,\mathrm{kHz}$ in e).
        An overview of the spectrum is displayed in e). 
        In a)-d), excerpts highlighting the actual Bloch--Siegert shift for different ratios $\omega_0/\omega_1$ 
        are shown. 
        The vertical purple dashed line corresponds to the peak predicted by FVV. 
        We observe that for $\omega_0\lesssim 3\omega_1$ the third order is essential 
        for predicting the correct location of the resonance, 
        which can be viewed at in f). 
    }
    \label{fig:BlochSiegert}
\end{figure*}
Fig.~\ref{fig:BlochSiegert} displays the corresponding spectra in second and third order
where we analyze the accuracy of the truncated series.
We are aware of the fact that $\omega_1/ \omega_0\ll 1$ is typical of experiments,
but we also study the range for $\omega_1/ \omega_0\lessapprox 1$ 
to gain insight at which point the approximation fails. 

The line shapes coincide 
for sufficiently small $\omega_1/ \omega_0$. 
It is clearly visible that neglecting the kick operator 
in FVV leads in this case to a flip in sign in the spectrum. 
For $\omega_1/ \omega_0 \propto 1/3$ in second order, a clear deviation between the perturbative approach and the numerically exact solution 
becomes visible. 
This error is corrected by adding the third order. 
The third-order effective Hamiltonian in FVV 
$F^{(3)}_\text{FVV} = -\frac{\omega_1^3}{16\omega_0^2}S^x$ 
predicts resonances to be centered at 
$\omega = 0,\,\pm\omega_1\sqrt{1-\frac{\omega_1^2}{16\omega_0^2}+\frac{\omega_1^4}{256\omega_0^4}}$. 
This essentially reproduces the numerically exact benchmark, 
as can be seen in Fig.~\ref{fig:BlochSiegert}. The SI Sect.~S2 shows in addition the behavior of the other two spin components. 
The Bloch--Siegert shift clearly illustrates that the kick operator has to be included into the calculation of the time evolution in the FVV scheme to obtain the correct signal.

\subsection{Magic-angle spinning spectra of dipolar-coupled spin systems}
\label{sub:MAS}

The second example is an application in which the periodic time dependence is due to a fast mechanical rotation of the sample. 
Magic-angle spinning 
\cite{andrewNuclearMagneticResonance1958,andrewRemovalDipolarBroadening1959,loweFreeInductionDecays1959,mehringPrinciplesHighResolution2012} 
is a key technique in solid-state NMR that 
partially eliminates detrimental line broadening caused by anisotropic interactions in solid samples. 
The rotation axis is tilted by the magic angle ($\theta_\text{m}\approx 54.74^\circ$) with respect to the 
static magnetic field, which defines the $z$-direction. 
Typically, the Hamiltonian is transformed into the Zeeman rotating frame,
and so-called non-secular terms, which oscillate fast with multiples of the Larmor frequency, are neglected. 
This is referred to as the secular approximation, and corresponding 
higher-order terms scale with the inverse Larmor frequency, 
which is usually much larger than the typical NMR interactions in spin-1/2 systems. 
The secular approximation  in NMR 
is equivalent 
to first-order FVV or AHT with respect to the inverse Larmor frequency.

\begin{figure*}
    \centering 
    \includegraphics[width=0.7\textwidth]{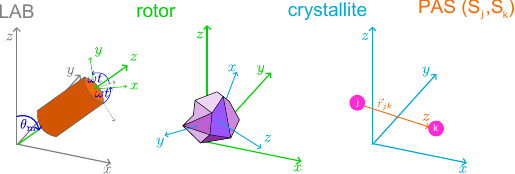}
    \caption{
        The rotation from the Principal-Axis System (PAS) of spin pair $jk$ to the lab frame is composed of three subsequent rotations. 
        1) The laboratory frame system (LAB) is rotated around the $y$-axis at the magic angle $\theta_\text{m}$, followed by a rotation around the new $z$-axis by $\omega_\text{r} t$. 
        2) The orientation of the crystallite-fixed frame in the rotor-fixed basis is defined by three Euler angles $(\alpha, \beta, \gamma)$. 
        3) Due to its axial symmetry, the PAS of $jk$ can be fully described by two angles $(\theta_{jk},\varphi_{jk})$ with regard to the crystallite-fixed frame. 
        These are the spherical coordinates of the connecting vector $\vec{r}_{jk}$, which provides the direction of the $z$-axis of the PAS of $jk$. 
    }
    \label{fig:lab_jk_rotation}
\end{figure*}

The mechanical rotation frequency of MAS can be up to about $200\,\mathrm{kHz}$, which is fast but not as large as the Larmor frequency. 
Thus, MAS represents a suitable use case for perturbative approaches, i.e., 
the computation of higher-order contributions in the inverse MAS frequency $\omega_\text{r}$. 
We consider a standard MAS NMR Hamiltonian including chemical shifts and homonuclear dipolar interaction 
in the Zeeman rotating frame in the secular approximation 
\begin{subequations}\label{eqn:MASHamiltonian}
    \begin{gather}
        H(t) = \sum_{i=1}^N \omega_i S_i^z  
        +\frac{1}{2} \sum_{i,j=1}^N \sum_{\substack{k=-2\\
                  (k\neq 0)} }^{2} \omega_{ij}^{(k)} e^{ik\omega_\text{r} t} \{ 3S_i^zS_j^z - \vec{S}_i\cdot \vec{S}_j \} \,, \\
        \omega_{ij}^{(k)} = \frac{1}{2} \delta_{ij} d^2_{k0}(-\theta_\text{m}) 
        \sum_{l=-2}^{2} e^{-il\alpha} d^2_{lk}(\beta) e^{-ik\gamma} 
        d^2_{0l}(-\theta_{ij}) e^{il\phi_{ij}} \,, \quad
        \delta_{ij} = -2\frac{\mu_0\gamma_i\gamma_j\hbar}{2\pi r_{ij}^3}\,. 
    \end{gather}
\end{subequations}
The anisotropy of the dipolar coupling $\delta_{ij}$ (we set $\delta_{ii}\equiv 0$) defines the strength of the dipolar interaction and 
determines the time-dependent pre-factor $\sum_k \omega_{ij}^{(k)}e^{ik\omega_\text{r} t}$.  
The full dipolar coupling  in the laboratory frame 
is defined by three subsequent rotations, 
which are illustrated in Fig.~\ref{fig:lab_jk_rotation}. 
The first transformation is from the principal axis system (PAS) of spin pair $(S_i,S_j)$ to the crystallite-fixed frame, 
often defined by the PAS of spin pair $(S_1,S_2)$. 
Then, the crystallite-fixed frame is rotated to the rotor-fixed frame with the $z$-axis along the rotation axis. 
The representation in the laboratory frame is obtained by the last rotation by the 
magic angle $\theta_\text{m}$ and the time-dependent rotor angle $\omega_\text{r} t$. 
The inter-nuclei vector $\vec{r}_{ij}$ is defined in the PAS of $(S_1,S_2)$ 
and possesses the spherical coordinates $(r_{ij}, \theta_{ij}, \phi_{ij})$. 
The Euler angles $(\alpha,\beta,\gamma)$ follow the $zyz$ convention and transform the PAS of $(S_1,S_2)$ 
to the rotor-fixed frame by three subsequent rotations $R^z(\alpha)$, $R^{y'}(\beta)$, $R^{z''}(\gamma)$. 
The reduced Wigner rotation matrix elements for rank-2 tensors~\cite{BrinkSatchler1993,mehringPrinciplesHighResolution2012}
are denoted by 
$d^2_{nm}(\beta)$ 
and are listed in the SI Sect.~S3. 

\begin{figure}
    \centering
    \includegraphics[height=3.5cm]{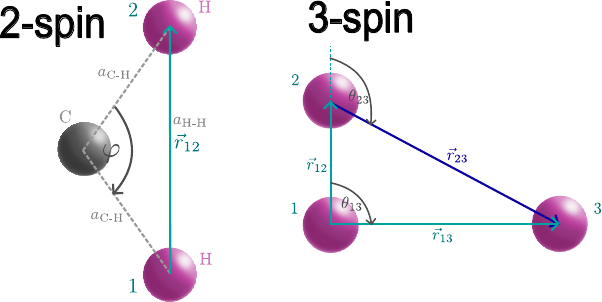}
    \caption{
        The two-spin (left panel) and three-spin (right panel) crystallites under MAS are considered, as a mimic of protons, e.g., within an alkane chain. 
    }
    \label{fig:2_3spinsystem}
\end{figure}

In the following, we consider crystallites with $N=2$ and $N=3$ spins, respectively, 
each as a single crystal with fixed orientation in the rotor and as a powder sample, 
for which the Zaremba-Conroy-Wolfsberg (ZCW) averaging scheme~\cite{chengInvestigationsNonrandomNumerical1973,Ponti:1999td} 
is applied. 
Technical details concerning the latter are summarized in the SI Sect.~S4. 
The initial density operator is $\rho = \sum_i S_i^x$ – just as for the Bloch–Siegert shift, we removed the part in $\rho$ that is proportional to the identity matrix and a potential prefactor –
and the detection operator is $\mathcal{S} = \sum_i S_i^+$. 
In the following, we provide some 
representative examples of our investigations.

\subsubsection{Two-spin system}

The system, illustrated in Fig.~\ref{fig:2_3spinsystem}, mimics the protons of a $\text{CH}_2$ group. 
We set the carbon–hydrogen bond length to $a_\text{C-H}=1.09\cdot 10^{-10}\,\mathrm{m}$ 
and assume ideal tetrahedral geometry, i.e., a bond angle of $\varphi = \cos^{-1}(-1/3)\approx 109.47^\circ$. 
Consequently, the distance between the two homonuclear dipolar coupled protons is set to $r_{12}=2a_\text{C-H}\sin(\varphi / 2) = 1.78\cdot 10^{-10}\,\mathrm{m}$. 
Using the gyromagnetic ratio of protons, $\gamma_1 = \gamma_2 = 2.675 \cdot 10^8 \,\mathrm{s}^{-1} \mathrm{T}^{-1}$, 
the anisotropy is 
$\delta_{12}/(2\pi) = -42.6\,\mathrm{kHz}$.
We investigated several crystal orientations and spinning frequencies. 
The chemical shifts are set to $\omega_1/(2\pi)=1\,\mathrm{kHz}$
and $\omega_2/(2\pi)=-1\,\mathrm{kHz}$,  
and the powder-averaged line shape comprises 
more than 1000 orientational samples. 
We present spectra for two different spinning frequencies~$\omega_\text{r}$ 
in order to illustrate the behavior in dependence of the perturbation parameter $\propto \rho(H) /\omega_\text{r}$. 

\begin{figure*}
    \includegraphics[width=0.99\textwidth]{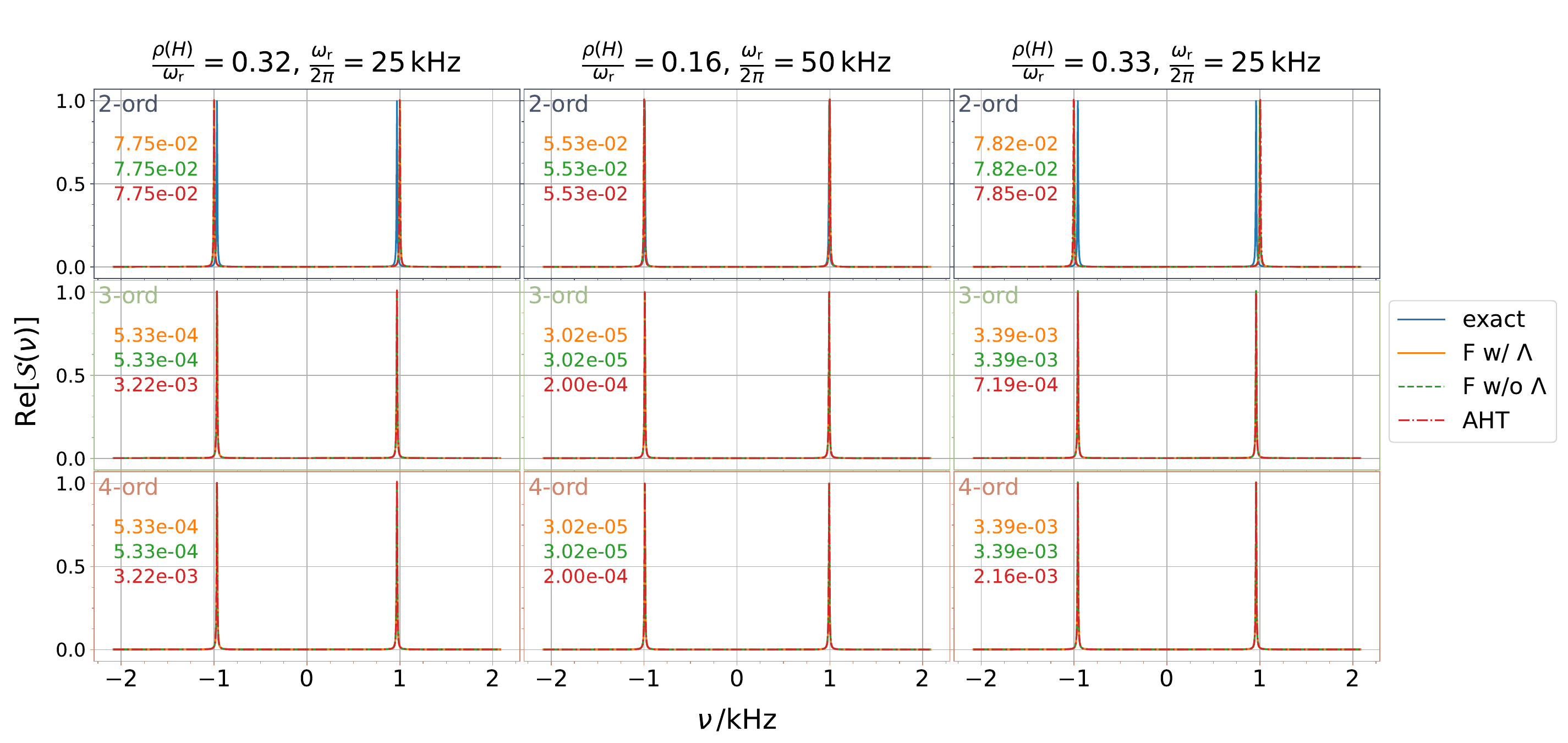}
    \centering
    \caption{
        Spectra of the dipolar-coupled two-spin system under MAS for two different crystal orientations and two different spinning frequencies. 
        The anisotropy of the dipolar coupling is $\delta_{12}/(2\pi)=-42.6\,\mathrm{kHz}$, 
        and the chemical shifts are $\omega_1/(2\pi) = 1\,\mathrm{kHz}$ and $\omega_2/(2\pi) = -1\,\mathrm{kHz}$. 
        The line broadening parameter is $\eta/(2\pi)=4.17\,\mathrm{Hz}$.
        The Euler angles of the crystallite's orientation with respect to the rotor are $(\alpha,\beta,\gamma)= (0, 0.2\pi, 0)$ in the 
        \textbf{left} and in the \textbf{middle} column, 
        and $(0, 0.4\pi, 0.3\pi)$ in the \textbf{right} column of panels. 
        This allows one to recognize differences in how FVV and AHT converge depending on the orientation. 
        The square root of the mean squared error is given by the colored numbers in each subplot. 
        The differences between FVV with kick operator (F w/ $\Lambda$) and without kick operator (F w/o $\Lambda$) are negligible. 
        For the majority of orientations, AHT yields a larger error than FVV. 
    }
    \label{fig:MAS2_crystal}
\end{figure*}

\begin{figure*}
    \includegraphics[width=0.73\textwidth]{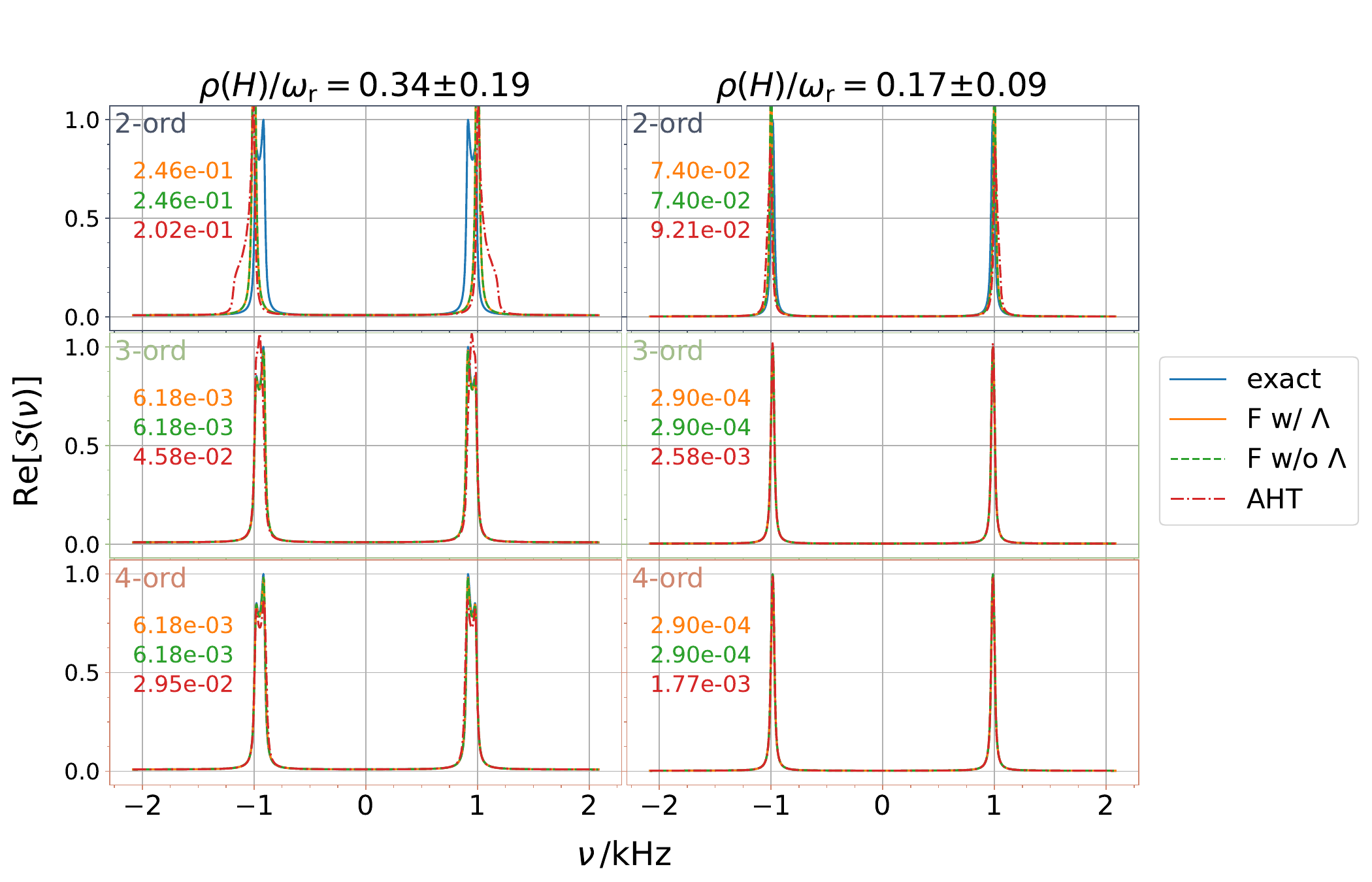}
    \centering
    \caption{
        The powder-averaged spectra of the dipolar-coupled two-spin system under MAS with two different rotation frequencies, 
        $\omega_\text{r}/(2\pi)=25\,\mathrm{kHz}$ on the left and $\omega_\text{r}/(2\pi)=50\,\mathrm{kHz}$ on the right. 
        The line broadening parameter is $\eta/(2\pi)=12.5\,\mathrm{Hz}$.
        The powder average is done over 1597 orientations following the ZCW scheme. 
        The square root of the mean squared error is given by the colored numbers in each subplot. 
        Differences between FVV with and without kick operator are negligible, and
        AHT yields a larger error than FVV. 
    }
    \label{fig:MAS2_powder}
\end{figure*}

Figure~\ref{fig:MAS2_crystal} shows the spectrum of a single crystal with a fixed orientation. 
We observe that the kick operator in the FVV approach does not make a visible difference in the spectra. 
The numerical values are not exactly the same, but the difference is very small. 
Furthermore, we recognize that FVV is the more accurate approximation of the numerically exact line shape compared to AHT for most 
orientations. 
To validate this observation, 
the line shape of the corresponding powder is provided in Fig.~\ref{fig:MAS2_powder}. 
Again, we note that AHT is less numerically robust than FVV
and converges more slowly. 
The difference between FVV with and without the kick operator exists, but appears only in small details of the line shape.

\subsubsection{Three-spin system}

\begin{figure*}
    \includegraphics[width=0.99\textwidth]{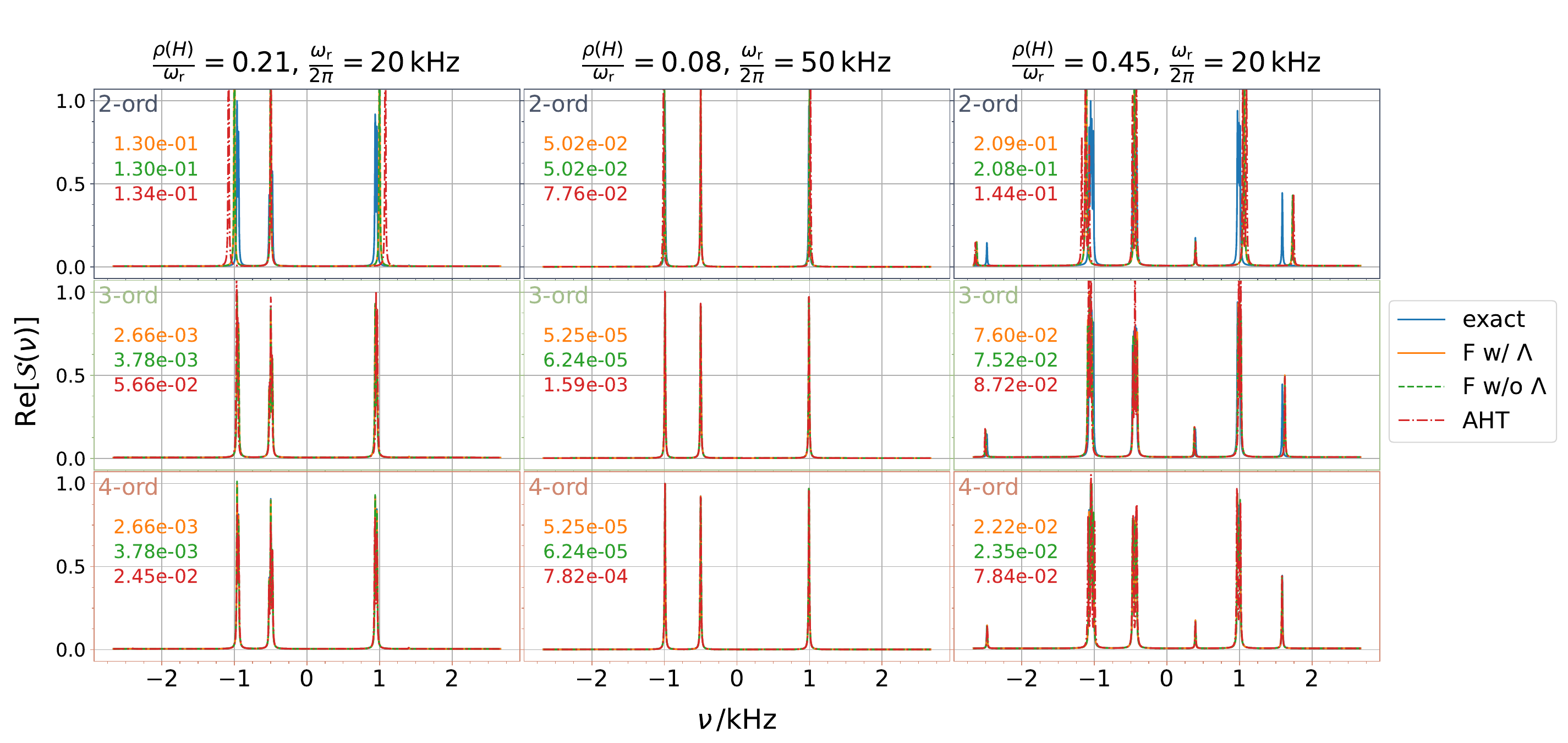}
    \centering
    \caption{
        Spectra of the dipolar-coupled three-spin system under MAS for two different orientations of the crystal and two different spinning frequencies.  
        The anisotropies of the dipolar couplings are $\delta_{12}/(2\pi)=-42.6\,\mathrm{kHz}$, 
        $\delta_{13}=0.155\delta_{12}$
        and $\delta_{23}=0.106\delta_{12}$, 
        and the chemical shifts are $\omega_1/(2\pi) = 1\,\mathrm{kHz}$, $\omega_2/(2\pi) = -1\,\mathrm{kHz}$ and $\omega_3/(2\pi) = 0.5\,\mathrm{kHz}$. 
        The spins within the crystallite are arranged with $\theta_{13}=\pi/2$ and $\theta_{23}=0.657\pi$. 
        The line broadening parameter is $\eta/(2\pi)=5.33\,\mathrm{Hz}$.
        The Euler angles of the crystallite with respect to the rotor are $(\alpha,\beta,\gamma)= (0., 0.1\pi, 0.4\pi)$ 
        in the \textbf{left} column and in the \textbf{middle} column, 
        and $(0.8\pi, 0.4\pi, 0.3\pi)$ in the \textbf{right} column of panels. 
        This allows one to recognize differences of FVV and AHT concerning their convergence depending on the orientation. 
        The square root of the mean squared error is given by the colored values. 
        A difference between FVV with kick operator (F w/ $\Lambda$) and without kick operator (F w/o $\Lambda$) becomes visible depending on the orientation of the crystallite. 
        AHT yields the largest error for most orientations. 
    }
    \label{fig:MAS3_crystal}
\end{figure*}

\begin{figure*}
    \includegraphics[width=0.73\textwidth]{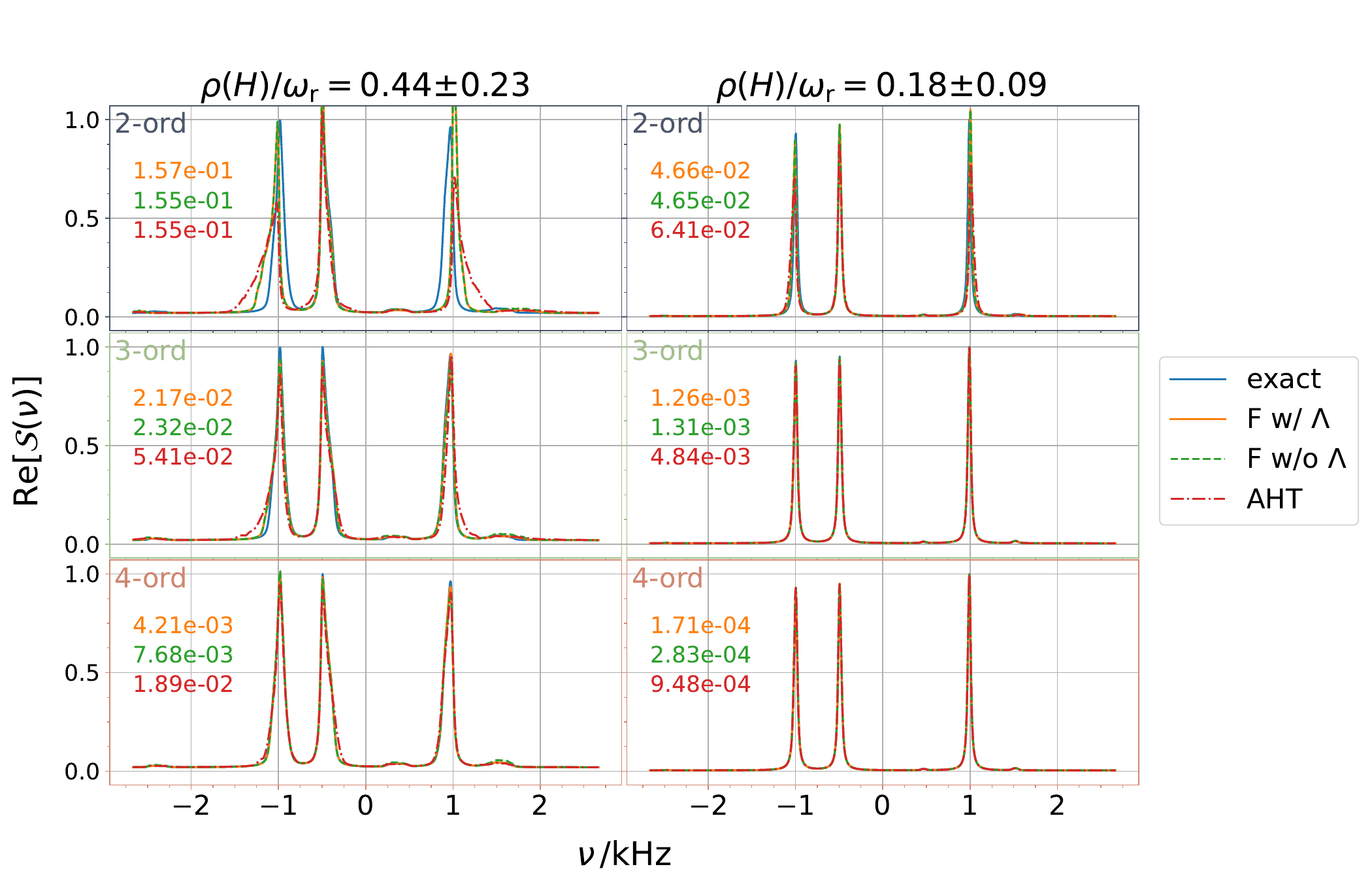}
    \centering
    \caption{
        Powder-averaged spectra of the dipolar-coupled three-spin system under MAS with two different rotation frequencies, 
        $\omega_\text{r}/(2\pi)=20\,\mathrm{kHz}$ on the left and 
        $\omega_\text{r}/(2\pi)=50\,\mathrm{kHz}$ on the right. 
        The line broadening parameter is $\eta/(2\pi)=1.6\,\mathrm{Hz}$.
        The powder average is done over 1154 
        orientations following the ZCW scheme. 
        The square root of the mean squared error is given by the colored values. 
        Differences between FVV with and without kick operator $\Lambda$ are discernibly, and 
        AHT implies the largest deviation. 
    }
    \label{fig:MAS3_powder}
\end{figure*}
The two-spin system from the previous paragraph is extended by a third spin, 
visualizing an additional, more distant proton. 
The assumed inter-nuclei vectors are $\vec{r}_{12} = (0,0,a_\text{H-H})^T$ 
and $\vec{r}_{13} = (1.86a_\text{H-H},0,0)^T$, as illustrated in Fig.~\ref{fig:2_3spinsystem}. 
This fixes the angles to 
$\theta_{13}=\pi/2$, $\theta_{23}=0.657\pi$ and $\phi_{13}=\phi_{23}=0$, 
and the anisotropies are related via 
$\delta_{13}=0.155\delta_{12}$
and $\delta_{23}=0.106\delta_{12}$ with $\delta_{12}/(2\pi)=-42.6\,\mathrm{kHz}$. 
Again, we present spectra for two different rotation frequencies, 
both for two fixed crystal orientations in Fig.~\ref{fig:MAS3_crystal}
and for a powder average in Fig.~\ref{fig:MAS3_powder}. 

Our observations are similar to those of the two-spin system. 
We clearly see that FVV is the more accurate approximation of the numerically exact line shape compared to AHT. 
The location of resonances in single crystal spectra shows faster convergence, 
and the general powder line shape is more quickly and more accurately approximated by FVV. 
Because the eigenvalues of the effective Hamiltonian $F_\text{FVV}^{[n]}$ and the stroboscopic Hamiltonian 
$\tilde{F}_\text{FVV}^{[n]}$ are exactly the same in each order $n$, 
the removal of the kick operator in FVV has no effect on the first moment of the resonances. 
However, in contrast to the two-spin system, 
the intensity of the peaks changes slightly
when the kick operator is added or removed. 

Further investigations of ours on the influence of the kick operator on the spectrum deduced by stroboscopic measurements show that in most cases 
$\Lambda_\text{FVV}(t_0)$ seems to be negligible for a Hamiltonian of the type given in Eq.~(\ref{eqn:MASHamiltonian}). 
The Eq.~(\ref{eqn:spectrum_summary}) for the calculation of the spectrum reveals that differences caused by the kick operator 
occur only in $\rho_{jk} \mathcal{S}_{kj}$. 
More precisely, the difference between the spectrum with 
and without the kick operator can be expressed by 
\begin{subequations}
\label{eq:difference}
    \begin{align}
        \Delta \mathcal{S}(\omega)
        &:= \sum_{jk} ( \tilde{\rho}_{jk} \tilde{\mathcal{S}}_{kj} - \rho_{jk} \mathcal{S}_{kj} ) \frac{1}{1 - e^{-\eta \cdot T} e^{i (\omega -\Delta \lambda_{jk})\cdot T}} \,, \\ 
        \tilde{\rho} & := e^{+i\Lambda(t_0)} \rho e^{-i\Lambda(t_0)}\,,\quad 
        \tilde{\mathcal{S}} := e^{+i\Lambda(t_0)} \mathcal{S} e^{-i\Lambda(t_0)}\,. 
    \end{align}
\end{subequations}
On the one hand, we find clear differences of $10-40\,\%$ in the operator norm 
between $S^{\alpha}:=  \sum_i S_i^\alpha$ and 
$\tilde{S}^{\alpha}:=  e^{+i\Lambda(t_0)} S^\alpha e^{-i\Lambda(t_0)}$ with $\alpha=x,y,z,\pm$ for both generic NMR Hamiltonians and for entirely random Hermitian matrices as Hamiltonian. 
On the other hand, omitting the kick operator in case of completely random Hermitian matrices 
leads to a difference in the corresponding spectrum that is clearly larger than what we
found in the case of MAS. 
We conclude that the differences of the measured 
operators are weighted in Eq.~(\ref{eq:difference}) 
in a way that does not significantly alter the actual spectra.
Thus, the kick operator is not very relevant to the FVV-based
calculation of the spectrum in Eq.~(\ref{eqn:spectrum_summary}). 
Supplementary investigations on this issue can be found in App.~\ref{app:kick_operator}.

\subsection{Non-stroboscopic measurements in MAS} 

Although the temporal discretization to stroboscopic time points is mathematically very convenient, it does not reflect common experimental practice.  
There, either non-rotor synchronized sampling is used, or additional measurements are taken in-between integer multiples of the period $T$, i.e., the timestep between observations is decreased from period $T$ to time increment $\tau=T/\alpha$ with integer $\alpha\geq 2$. 
For the latter case, the expression of the spectrum in Eq.~(\ref{eqn:spectrum_summary})  
needs to be adapted in order to take the intra-stroboscopic dynamics into account, as well. 
By separating the stroboscopic measurements from the intra-stroboscopic ones,   
one arrives at 
\begin{subequations}
    \begin{align}
        \mathcal{S}(\omega) &= \frac{1}{\alpha} \sum_{n=0}^\infty \sum_{k=0}^{\alpha-1} e^{i\omega (nT+k\tau)} \mathcal{S}(nT+k\tau) \\
        &= \sum_{ijlm} \Psi_{ijlm} \rho_{jl} \mathcal{S}_{mi} \frac{1}{1 - e^{-\eta \cdot T} e^{i (\omega -\Delta \lambda_{jl})\cdot T}} \\ 
        \Psi_{ijlm} &:= \frac{1}{\alpha} \sum_{k=0}^{\alpha-1} e^{i\omega k\tau} \bigl(U(t_0+k\tau,t_0)\bigr)_{ij} \bigl(U^\dagger(t_0+k\tau,t_0)\bigr)_{lm}
    \end{align}
\end{subequations}
as a formula for the signal in frequency space. 
Again, the indices of $\rho$, $\mathcal{S}$, and $U(t_0+k\tau,t_0)$ refer to the matrix entries in the eigenbasis of the single-cycle evolution operator $U(t_0+T,t_0)$. 
A detailed derivation is provided in the SI Sect.~S1. 

This third example addresses two main questions. 
First, how does the intensity of the central band change in case of additional intra-stroboscopic measurements? 
Second, how good is the convergence of AHT and FVV in the modeling of the side bands? 
According to FME, 
the kick operator is explicitly needed in order to access intra-stroboscopic dynamics.
Apart from the numerically exact benchmark, we consider spectra that are computed by 
\begin{enumerate}
    \item[a)] AHT $e^{-i\Lambda(t)}e^{-iF\cdot (t-t_0)}$, 
    \item[b)] FVV $e^{-i\Lambda(t)}e^{-iF\cdot (t-t_0)}e^{i\Lambda(t_0)}$.  
\end{enumerate}

We stick to the Hamiltonian in Eq.~(\ref{eqn:MASHamiltonian}) for the three spins that are 
illustrated in Fig.~\ref{fig:2_3spinsystem}. 
The values for $\omega_{1,2,3}$, $\delta_{12,13,23}$, $\theta_{12,13,23}$, and $\phi_{12,13,23}$ remain the same, 
as well as the initial density operator $\rho$ and the detection operator $\mathcal{S}$.  
The spinning frequency is set to $\omega_\text{r}/(2\pi) = 25\,\mathrm{kHz}$ resulting in $\rho(H)/\omega_\text{r}=0.28$. 
A single crystal orientation $(\alpha,\beta,\gamma)=(0.2\pi, 0.3\pi, 0.7\pi)$ is used 
because it gives more insightful results 
than the powder average over all orientations. 
The line broadening parameter is set to $\eta/(2\pi)=16\,\mathrm{Hz}$.

\subsubsection{Intensity of the central band}

\begin{figure*}
    \includegraphics[width=0.75\textwidth]{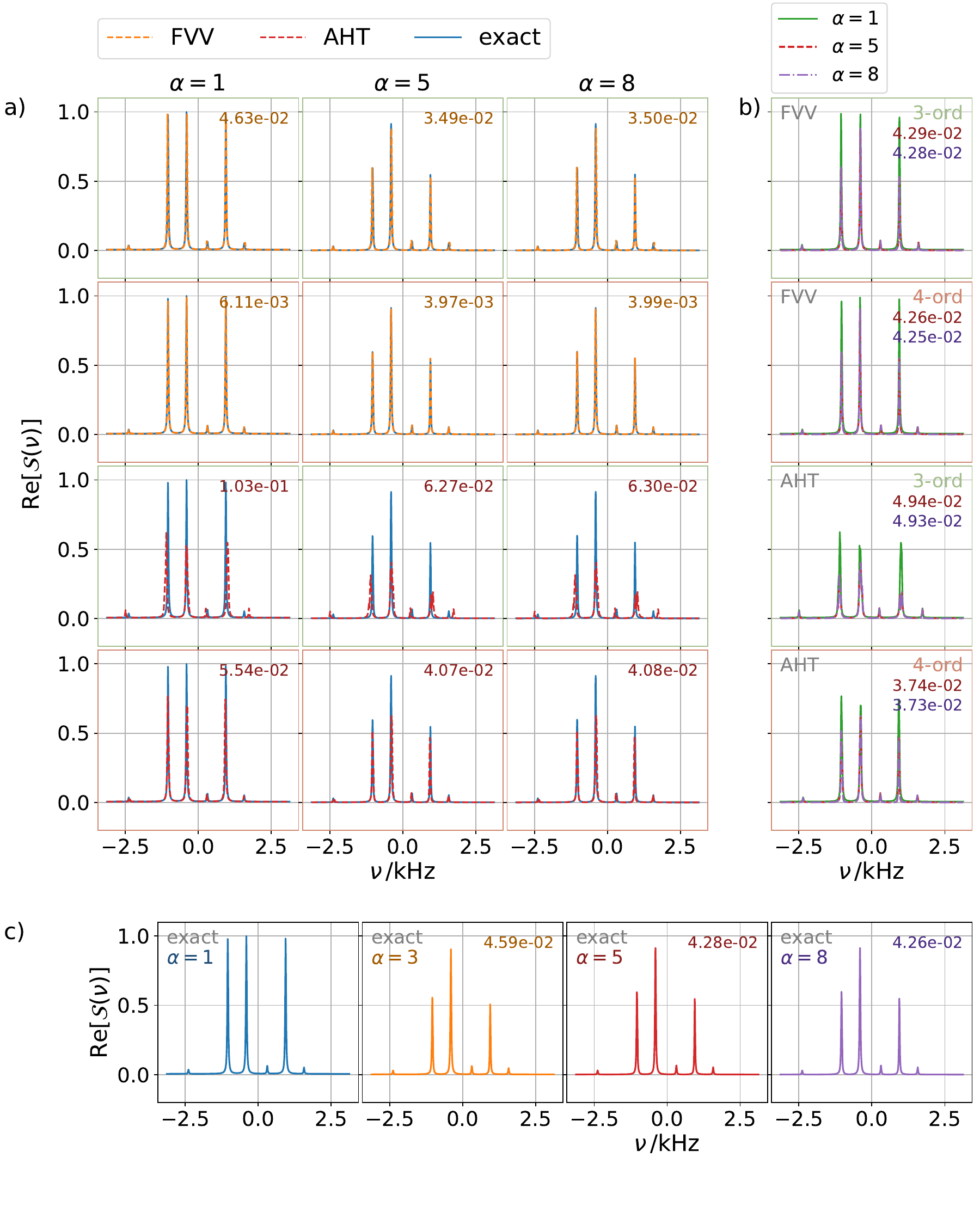}
    \centering
    \caption{
        Behavior of the central band of a three-spin crystal under MAS when the number of intra-stroboscopic measurements $\alpha$ is varied. 
        The dynamics are sampled at $t=n\tau$ with $\tau=T/\alpha$, $n\in\mathbb{N}^0$, at spinning frequency $\omega_\text{r}/(2\pi)=25\,\mathrm{kHz}$. 
        The upper left figure a) compares FVV (first and second row) and AHT (third and fourth row) to the numerically exact line shape. 
        The deviation from the numerically exact line shape is measured using the square root of the mean squared error, and is given by the colored numbers. 
        The right column b) presents the influence of $\alpha$ within FVV or AHT, 
        the numbers are the deviation between $\alpha=1$ and $\alpha\neq 1$. 
        The bottom row c) reflects the numerically exact pendant to b), i.e., 
        the values represent the deviation between $\alpha=1$ and $\alpha\neq 1$. 
    }
    \label{fig:centerband}
\end{figure*}

Figure~\ref{fig:centerband} provides information on the intensities of the central band 
using intra-stroboscopic sampling. 
First, we note from the bottom panel that the central band clearly changes if the signal is sampled at additional intra-stroboscopic time points. 
More precisely, the location of the resonances is unchanged, but
neither their absolute weight 
nor the ratio between the amplitudes of the resonances stays the same. 
This is also the case for the perturbative schemes AHT and FVV illustrated in the right panel. 

Moreover, considering the upper left panel of Fig.~\ref{fig:centerband}, FVV's advantage relative to AHT is apparent once more. 
The results obtained by FVV unequivocally match the full line shape in general and the resonances in particular better than the ones obtained by AHT.

We conclude for both the exact and the perturbative approaches that the first moments of resonances are independent of the sampling frequency.
However, the precise intensities as well as the ratios between each other are
significantly influenced by additional intra-stroboscopic time points. 

\begin{figure*}
    \includegraphics[width=0.8\textwidth]{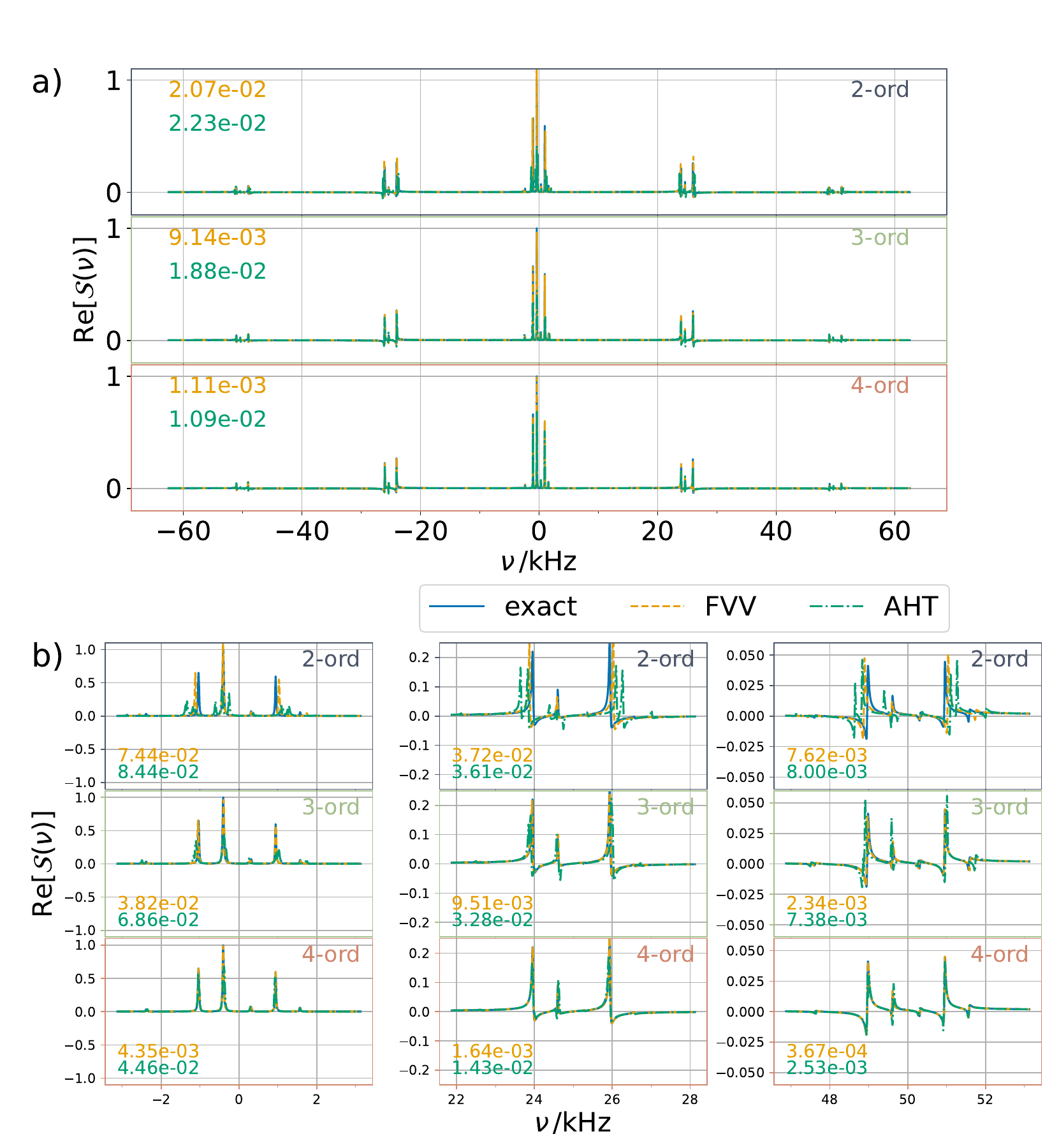}
    \centering
    \caption{
        The full spectrum resulting from additional intra-stroboscopic time points at $t=n\tau$ with $\tau=T/5$, $n\in\mathbb{N}^0$, 
        of a three-spin crystal under MAS. 
        The upper figure a) presents the full spectrum $\nu\in [-1/(2\tau), 1/(2\tau)]$ reflecting that the kick operator in both approaches (AHT and FVV)
        is capable of and crucial for modeling the side bands, which arise 
        at integer multiples of the rotation frequency $\omega_\text{r}/(2\pi)=25\,\mathrm{kHz}$.
        The numbers are the square root of the mean squared error 
        and measure the deviation between the approximate scheme and the numerically exact line shape. 
        Figure b) provides a close-up of the center line (left column) and of the side bands (central and right column) for positive~$\nu$. 
        Equally to a), the values quantify the deviation of FVV and AHT from the exact spectrum in this specific excerpt. 
    }
    \label{fig:sidespectra}
\end{figure*}

\subsubsection{Convergence of side bands}

In Fig.~\ref{fig:sidespectra}, the spectrum is given for a time increment $\tau=T/5$, both 
in a full overview and in close-ups for the central and side bands. 
First, we want to emphasize that the kick operator is crucial in order to make the side bands visible at all. 
Neglecting the kick operator would remove the micromotion 
within each period and thereby erase the contributions at
higher frequencies. 
Apart from that, both the computed errors for FVV and AHT and the 
performance in the presented spectra corroborate the conclusion 
that FVV is more robust and accurate than AHT. 
FVV matches the location of resonances better, its mean squared error in the spectrum is obviously smaller,  and the convergence is faster. 
We highlight that the fourth-order result of AHT is still worse than the third-order result
of FVV.

\section{Conclusion and outlook}
\label{sec:summary}

Most solid-state NMR experiments involve a periodic time-dependent Hamiltonian with different time and energy scales, which is exploited for their analysis. 
This leads to the concept of effective Hamiltonians that yield effective time-independent 
interactions induced by the temporal modulation. 
Both Average Hamiltonian Theory (AHT) and Floquet theory combined with Van Vleck perturbation theory (FVV) are widely used in the NMR community.
Yet, it is not always clear why a particular approach is chosen. 

This issue is addressed in our work by an extensive analysis with emphasis on NMR-relevant applications. We recall how AHT and FVV are derived
in the more general Floquet--Magnus Expansion (FME), 
which allows one to identify clear relations between AHT and FVV expressions.
In particular, it proves their perturbative equivalence. 
Their apparent differences are entirely 
explained by the truncation of the infinite series expansion. 
The latter is a subtle, but important point because our calculations show that FVV is more robust, reliable and efficient than AHT. This holds both for the  
approximation of the temporal unitary propagator and  for the  NMR spectra in frequency space. 
We conclude that the effective Hamiltonian of Floquet theory 
combined with the kick operator is the best generic choice  for solid-state NMR. 


On the technical side, we found that the derivation of higher orders via integration is prone to mistakes which is why we suggest an alternative calculation scheme that does not require explicit integration thereby avoiding to distinguish proliferating cases.
This helped us to provide the first four orders of both FVV and AHT. 
Our results reveal the significance of the kick operator. 
We apply AHT and FVV to calculate NMR spectra in two different system classes: 
the Bloch--Siegert shift and homonuclear dipolar coupled spin systems under magic-angle spinning. Furthermore,
we showcase how additional intra-stroboscopic sampling enables the computation of sidebands 
and how this influences the shape of the central band. 
For this, the kick operator is crucial. 
The obtained results lead us to recommend the use of the FVV 
approach for future analyses of NMR applications. 

There are still open questions calling for further 
methodological research. 
One such issue concerns the convergence mechanisms of FME in general, but especially FVV, which is discussed but
not yet fully solved with mathematical rigor.
Based on our exemplary calculations and numerical validation routines, 
we empirically found that FVV provides the superior framework for an approximation both 
of the temporal evolution operator and of the resulting spectra. 
Yet, this is not mathematically proved by bounding the remaining deviations.

Another question deals with the impact of the kick operator, 
which seems to be negligible in some NMR spectra for stroboscopic measurements.
While we could provide upper bounds for the deviations in the spectra without
kick operator these bounds are not very tight.
A comprehensive, rigorous mathematical exploration is called for.


\section*{Code and data availability}
All data used in the paper are contained in the publication and the SI and can be
generated using the equations given there.

\section*{Author contributions}
GSU and ME designed the research. AJB did all calculations, simulations, and the initial writing of the paper. The simulations were partially counter-validated by and discussed with ME.  
ME did the Mathematica calculation in the SI. 
All authors discussed the results and edited, reviewed and approved the final manuscript.

\section*{Competing interests}
ME is executive editor of Magnetic Resonance. 
The authors have no other competing interests to declare.

\section*{Acknowledgements}
AJB acknowledges a fellowship from the Studienstiftung des deutschen Volkes.
We thank A. Eckardt for insightful discussions. 

\section*{Supplementary Information}
Supplementary Information is available as an ancillary file with this arXiv submission.

\clearpage
\noindent
\textbf{\LARGE Appendix}

\appendix


\section{Calculation scheme: Solving FME without integration}
\label{app:calc_scheme}

The following shows how higher-order terms in FME are derived in the Fourier basis. 
We believe that this approach has advantages because multidimensional integration is not necessary. 

First, the Fourier series of $\Lambda(t)=\sum_m\Lambda_{m}e^{im\omega t}$ and $H(t)=\sum_mH_{m}e^{im\omega t}$ are inserted into the defining differential 
Eq.~(\ref{eqn:FME_dgl}). 
By induction, the Fourier coefficients of the right-hand side are identified and equated 
to those of the left-hand side. 
With the help of the short-hand notation for nested commutators 
\begin{equation}
    [A_1,A_2,A_3,\dots] := [A_1,[A_2,[A_3,[\dots]]]] \,, 
\end{equation}
this leads to 
\begin{equation}\label{eqn:FME_fourier}
    \begin{aligned}
        im\omega\cdot \Lambda_m
        &= -F\delta_{m,0} + H_m 
        + \frac{i}{2} \Bigl\{ \sum_{n_1} [\Lambda_{n_1}, H_{m-n_1}] 
        + [\Lambda_{m}, F]  \Bigr\} \\ 
        &+ \sum_{j=2}^\infty (-i)^j \frac{B_j}{j!} \Bigl\{ \sum_{n_1,\dots,n_j} [\Lambda_{n_1},\dots,\Lambda_{n_j}, H_{m-n_1-\dots-n_j}] \\ 
        &+ (-1)^{j+1} \sum_{n_1,\dots,n_{j-1}} [\Lambda_{n_1},\dots,\Lambda_{m-n_1-\dots-n_{j-1}} , F]  \Bigr\} \,. 
    \end{aligned}
\end{equation}
Considering $m=0$ and $m\neq 0$ separately, 
Eq.~(\ref{eqn:FME_fourier}) allows for explicit formulas of $F$ and $\Lambda_{m\neq 0}$ 
\begin{equation}
    \begin{aligned}
        F&= H_0 + \frac{i}{2} \Bigl\{ \sum_{n_1} [\Lambda_{n_1}, H_{-n_1}] 
        + [\Lambda_{0}, F]  \Bigr\} \\ 
        &+ \sum_{j=2}^\infty (-i)^j \frac{B_j}{j!} \Bigl\{ \sum_{n_1,\dots,n_j} [\Lambda_{n_1},\dots,\Lambda_{n_j}, H_{-n_1-\dots-n_j}] \\
        &+ (-1)^{j+1} \sum_{n_1,\dots,n_{j-1}} [\Lambda_{n_1},\dots,\Lambda_{-n_1-\dots-n_{j-1}} , F]  \Bigr\} \,, 
    \end{aligned}
\end{equation}
\begin{equation}
    \begin{aligned}
        \Lambda_{m\neq 0} &= \frac{H_m}{im\omega} 
        + \frac{1}{2m\omega} \Bigl\{ \sum_{n_1} [\Lambda_{n_1}, H_{m-n_1}] 
        + [\Lambda_{m}, F]  \Bigr\} \\ 
        &+ \sum_{j=2}^\infty (-i)^{j+1} \frac{B_j}{j!m\omega} \Bigl\{ \sum_{n_1,\dots,n_j} [\Lambda_{n_1},\dots,\Lambda_{n_j}, H_{m-n_1-\dots-n_j}] \\
        &+ (-1)^{j+1} \sum_{n_1,\dots,n_{j-1}} [\Lambda_{n_1},\dots,\Lambda_{m-n_1-\dots-n_{j-1}} , F]  \Bigr\} \,. 
    \end{aligned}
\end{equation}

Finally, 
all relevant orders on both sides are equated. 
In the first orders, this leads to 
\begin{subequations}
    \begin{align}
        F^{(1)} &= H_0 \,, \\ 
        \Lambda_{m\neq 0}^{(1)} &= \frac{H_m}{im\omega} \,, \\ 
        F^{(2)} &= \frac{i}{2} \Bigl\{ \sum_{n_1} [\Lambda_{n_1}^{(1)}, H_{-n_1}]
        + [\Lambda_{0}^{(1)}, F^{(1)}] \Bigr\} \\ 
        &= \sum_{n\neq 0} \frac{[n, {-n}]}{2n\omega}
        + i\cdot [\Lambda_{0}^{(1)}, H_0] \,, \\ 
        \Lambda_{m\neq 0}^{(2)} &= \frac{1}{2m\omega} \Bigl\{
            \sum_{n_1} [\Lambda_{n_1}^{(1)}, H_{m-n_1}] 
            + [\Lambda_{m}^{(1)}, F^{(1)}]  
            \Bigr\} \\ 
            &= 
            \sum_{n\neq 0} \frac{[n, {m-n}] }{2inm\omega^2}
            + \frac{[m, 0]}{2im^2\omega^2} 
            + \frac{[\Lambda_{0}^{(1)}, H_{m}]}{2m\omega} \,. 
    \end{align}
\end{subequations}
In order to obtain the final expression, 
the gauge freedom in $\Lambda_{0}^{(n)}$ has to be fixed. 
This is especially simple for FVV where $\Lambda_{0}^{(n)}=0\,\forall n$. 
In AHT, it is $\Lambda_{0}^{(n)}=-\sum_{m\neq 0}\Lambda_m^{(n)}e^{im\omega t_0}$.

\section{Fourth-order FME}
\label{app:4thorder}

For the fourth-order single-cycle evolution operator $U^{[4]}(t_0+T,t_0)$, 
the fourth order effective Hamiltonian $F^{(4)}$ and the kick operator in third order $\Lambda^{(3)}$ are needed. 
Due to their length, they are not included in the main article but in the following appendix for both FVV and AHT. 
The fourth-order effective Hamiltonian of FVV agrees with the version published by \cite{aebischerLineWidthMagicAngle2024}. 
As far as we are aware, there is no fourth-order effective Hamiltonian available in commutator form in AHT. 
For completeness, we adapted and included the integral form from Ref.~\cite{blanesMagnusExpansionIts2009}, which is also part of our Mathematica validation calculation presented in the SI Sect.~S5. 

\subsection{Floquet--Van Vleck theory}

\begin{subequations}
    \begin{align}
        F^{(4)} &= 
        \sum_{n\neq 0} \sum_{\substack{k\neq 0 \\ (n+k\neq 0)}} \sum_{\substack{\ell\neq 0 \\ (n+k+\ell\neq 0)}} \Biggl\{ 
            \frac{\bigl[ [n,k], [\ell, -n-k-\ell] \bigr]}{24 (n+k+\ell)(n+k)n \omega^3} \\ 
            &+ \frac{\Bigl[ n, \bigl[ k, [\ell, -n-k-\ell] \bigr] \Bigr]}{24nk\ell \omega^3}
            + \frac{\Bigl[ -n-k-\ell, \bigl[ \ell, [n, k] \bigr] \Bigr]}{6 n (n+k) (n+k+\ell) \omega^3} 
            \Biggr\} \\ 
            &+ \sum_{n\neq 0} \sum_{\substack{k\neq 0 \\ (n+k\neq 0)}} \Biggl\{ 
                \frac{\bigl[ [-n-k,0], [n,k] \bigr]}{12 n (n+k)^2 \omega^3} \\ 
                &+ \frac{\Bigl[ -n-k, \bigl[ k, [n, 0] \bigr] \Bigr]}{3n^2 (n+k) \omega^3}
                + \frac{\Bigl[ -n-k, \bigl[ 0, [n, k] \bigr] \Bigr]}{4 n (n+k)^2 \omega^3}
                \Biggr\} \\ 
                &+ \sum_{n\neq 0} \sum_{k\neq 0} 
                \frac{\Bigl[ n, \bigl[ -n, [-k, k] \bigr] \Bigr]}{8 n^2 k \omega^3} 
                + \sum_{n\neq 0} 
                \frac{\Bigl[ -n, \bigl[ 0, [n, 0] \bigr] \Bigr]}{2 n^3 \omega^3} 
    \end{align}
\end{subequations}

\begin{subequations}
    \begin{align}
        \Lambda_{m\neq 0}^{(3)} &= 
        \frac{\bigl[0,[0,m]\bigr]}{im^3\omega^3} 
        + \sum_{n\neq 0} \frac{6n+m}{12im^2n^2\omega^3} \bigl[n,[m,-n]\bigr] \\
        &+ \sum_{\substack{n\neq 0 \\ (n\neq m)}} \frac{3m^2 -mn -3n^2}{6im^2n^2(m-n)\omega^3} \bigl[m-n,[0,n]\bigr] \\
        &+ \sum_{n\neq 0} \sum_{\substack{k\neq 0 \\ (n+k\neq 0)}} \frac{2k-n}{6inkm(n+k)\omega^3} \bigl[n,[k,m-n-k]\bigr] 
    \end{align}
\end{subequations}

\subsection{Average Hamiltonian theory}

\begin{subequations}
    \begin{align}
        F^{(4)} &= \frac{i}{12T} \int_{t_0}^{t_0+T}dt_1 \,\int_{t_0}^{t_1}dt_2 \, \int_{t_0}^{t_2}dt_3 \,\int_{t_0}^{t_3}dt_4 \times \\
        &\biggl\{ \Bigl[H(t_4), \bigl[H(t_3), [H(t_1), H(t_2)]]] 
         + \Bigl[H(t_1), \bigl[H(t_4), [H(t_3), H(t_2)]\bigr]\Bigr] \\ 
        &+ \Bigl[H(t_1), \bigl[H(t_2), [H(t_3), H(t_4)]\bigr]\Bigr] 
        + \Bigl[H(t_2), \bigl[H(t_3), [H(t_4), H(t_1)]\bigr]\Bigr]
        \biggr\}
    \end{align}
\end{subequations}

\begin{subequations}
    \begin{align}
        F^{(4)} &= 
        \sum_{n\neq 0} \sum_{\substack{k\neq 0 \\ (n+k\neq 0)}} \sum_{\substack{\ell\neq 0 \\ (n+k+\ell\neq 0)}} \Biggl\{ 
            \frac{\bigl[ [n,k], [\ell, -n-k-\ell] \bigr]}{24 (n+k+\ell)(n+k)n \omega^3} \\ 
            &+ \frac{\Bigl[ n, \bigl[ k, [\ell, -n-k-\ell] \bigr] \Bigr]}{24nk\ell \omega^3}
            + \frac{\Bigl[ -n-k-\ell, \bigl[ \ell, [n, k] \bigr] \Bigr]}{6 n (n+k) (n+k+\ell) \omega^3} 
            \Biggr\} \\ 
            &+\sum_{n\neq 0} \sum_{\substack{k\neq 0 \\ (n+k\neq 0)}} \sum_{\substack{m\neq 0 \\ (m-n-k\neq 0)}} \frac{e^{im\omega t_0}}{\omega^3} \Biggl\{ 
                \frac{2k-n}{6 n k m (n+k)} \Bigl[ 0, \bigl[ n, [k, m-n-k] \bigr] \Bigr] \\ 
                &+ \frac{\Bigl[ m-n-k, \bigl[ 0, [n, k] \bigr] \Bigr]}{4 (m-n-k) (n+k) k} 
                +\frac{\bigl[ [k,n], [m-n-k, 0] \bigr]}{4 k (n+k) (m-n-k)}  
                + \frac{\Bigl[ n, \bigl[ m-n-k, [0, k] \bigr] \Bigr]}{6 n (m-n-k) k} 
                \Biggr\} \\ 
                &+\sum_{n\neq 0} \sum_{k\neq 0} \sum_{m\neq 0} \frac{e^{im\omega t_0}}{\omega^3} 
                \frac{\Bigl[ m, \bigl[ n, [-n-k, k] \bigr] \Bigr]}{3 m n k} \\ 
                &+\sum_{n\neq 0} \sum_{k\neq 0} \sum_{\substack{m\neq 0 \\ (m-k\neq 0)}} \frac{e^{im\omega t_0}}{\omega^3} \Biggl\{
                    \frac{\Bigl[ m-k, \bigl[ k, [n, -n] \bigr] \Bigr]}{4 (m-k) k n}  
                    + \frac{\bigl[ [n, -n], [k, m-k] \bigr]}{4 nmk } 
                    \Biggr\} \\ 
                    &+\sum_{n\neq 0}\sum_{k\neq 0} \frac{3 \Bigl[ n, \bigl[ -n, [-k, k] \bigr] \Bigr]}{8 n^2 k \omega^3}  
                    +\sum_{n\neq 0}\sum_{\substack{k\neq 0 \\ (n+k\neq 0)}} \Biggl\{ 
                        \frac{\Bigl[ n, \bigl[ 0, [k, -n-k] \bigr] \Bigr]}{2 n^2 k \omega^3} \\  
                        &+ \frac{2n+k}{6 nk^2(n+k) \omega^3} \Bigl[ n, \bigl[ -n-k, [0, k] \bigr] \Bigr] 
                        +\frac{\bigl[ [-n-k, 0], [n, k] \bigr]}{3 n (n+k)^2 \omega^3 } 
                        \Biggr\} \\ 
                        &+ \sum_{n\neq 0}\sum_{m\neq 0} \frac{e^{im\omega t_0}}{\omega^3} \Biggl\{  
                            \frac{6n+m }{12 m^2 n^2} \Bigl[ 0, \bigl[ n, [m, -n] \bigr] \Bigr] \\ 
                            &+ \frac{\bigl[ [n, -n], [m, 0] \bigr]}{2nm^2 } 
                            +\frac{\Bigl[ m, \bigl[ n, [-n, 0] \bigr] \Bigr] + \Bigl[ n, \bigl[ m, [-n, 0] \bigr] \Bigr]}{6 mn^2 } 
                            \Biggr\} \\ 
                            &+ \sum_{n\neq 0}\sum_{m\neq 0} \frac{e^{i (n+m) \omega t_0}}{4nm^2\omega^3} \Biggl\{  
                                \Bigl[ n, \bigl[ 0, [0, m] \bigr] \Bigr] 
                                + \bigl[[m, 0], [n, 0] \bigr] 
                                \Biggr\} \\ 
                                &+ \sum_{n\neq 0}\sum_{\substack{m\neq 0 \\ (m-n\neq 0)}} \frac{e^{i m \omega t_0}}{\omega^3} \Biggl\{  
                                    \frac{4mn-2n^2-m^2}{2n(m-n)^2m^2} \Bigl[ 0, \bigl[ n, [0, m-n] \bigr] \Bigr] \\ 
                                    &+ \frac{\Bigl[ m-n, \bigl[ 0, [0, n] \bigr] \Bigr]}{4n^2(m-n)}
                                    + \frac{\bigl[[n, 0], [m-n, 0] \bigr]}{4n^2 (m-n)} 
                                    \Biggr\} \\     
                                    &+ \sum_{n\neq 0} \Biggl\{
                                        \frac{3\Bigl[ n, \bigl[ 0, [0, -n] \bigr] \Bigr]}{4n^3\omega^3} 
                                        + \frac{ \bigl[ [-n, 0], [n, 0] \bigr] }{4n^3\omega^3} 
                                        \Biggr\}  
                                        + \sum_{m\neq 0} 
                                        e^{im\omega t_0} \frac{\Bigl[ 0, \bigl[ 0, [0, m] \bigr] \Bigr]}{m^3\omega^3}
    \end{align}
\end{subequations}

\begin{subequations}
        \begin{align}
            \Lambda_{m\neq 0}^{(3)} &= 
            \frac{\bigl[0,[0,m]\bigr]}{im^3\omega^3} 
            + \sum_{n\neq 0} \bigl[n,[m,-n]\bigr] \frac{3n+m}{6im^2n^2\omega^3} 
            + \sum_{\substack{n\neq 0 \\ n\neq m}} \bigl[m-n,[0,n]\bigr] \frac{3m-n}{6in^2(m-n)m\omega^3} \\
            &+ \sum_{n\neq 0} \sum_{\substack{k\neq 0 \\ (n+k\neq 0)}} \bigl[n,[k,m-n-k]\bigr] \frac{2k-n}{6ink(n+k)m\omega^3} 
            + \sum_{n\neq 0} \frac{\bigl[m,[n,0]\bigr]}{2imn^2\omega^3} e^{in\omega t_0} \\
            &+ \sum_{n\neq 0} \sum_{k\neq 0} \frac{\bigl[k,[-n,m+n]\bigr]}{4imnk\omega^3} e^{ik\omega t_0} 
            + \sum_{\substack{n\neq 0 \\ (n+m\neq 0)}} \sum_{k\neq 0} \frac{\bigl[-n,[k,n+m]\bigr]}{12in(n+m)k\omega^3} e^{ik\omega t_0} \\ 
            &+ \sum_{n\neq 0} \sum_{\substack{k\neq 0 \\ (n-k\neq 0)}} \bigl[n,[k-n,m]\bigr] \frac{3n-k}{6inkm(k-n)\omega^3} e^{ik\omega t_0} 
            + \sum_{n\neq 0} \frac{\bigl[n,[0,m]\bigr]}{4inm^2\omega^3} e^{in\omega t_0}
        \end{align}
\end{subequations}

\section{Technical details of the numerical validation}
\label{app:toysystems}

We validated our formulas for FVV and AHT in Sect.~\ref{sub:numericalvalidation} for several systems with regard to 
their eigenvalues and eigenframe. 
The following provides details about the concrete systems we considered and how we measured the error. 

\subsection{Model} 

All considered systems are described by a Fourier series with a block-band structure, i.e., 
we define the time-periodic Hamiltonian 
\begin{equation}
    H(t) = \sum_{n=-C}^{C} H_n e^{in\omega t}\,, 
    \label{eqn:Hamiltonian}
\end{equation}
where the cutoff $C$ is fixed. 
In $d\in\mathbb{N}$ dimensions, $H(t)$ is a $d\times d$ Hermitian matrix, 
and we consider $d=2, 4, 8 , 16$, corresponding to a system of 1, 2, 3, 4 $S=1/2$-spins, respectively. 
We work in natural units, i.e., $\hbar=1$, and energies are measured in units of the spectral radius $\rho(H)$.

For $d=2$, the Hamiltonian is determined by 
\begin{equation}\label{eqn:toysystem}
    \begin{aligned}
        H(t) = J_0 \hat{S}_{\nu_0} + \sum_{n=1}^{C} \Bigl[s_n \hat{S}_{\mu_n} \sin(n \omega t) + c_n \hat{S}_{\nu_n} \cos(n \omega t)\Bigr]\,. 
    \end{aligned}
\end{equation}
The coefficients $J_0$, $s_n$, and $c_n$ are independently sampled from the uniform distribution $\mathcal{U}[-1, 1]$, 
while $\mu_n,\nu_n\in\{0,1,2,3\}$ are independently drawn from the discrete uniform distribution. 
We define $\hat{S}_0=\mathbb{1}$ and $\hat{S}_i=\sigma_i/2$ for $i=1,2,3$, with $\sigma_i$ denoting the Pauli matrices. 
We explored various cutoff parameters $C$ ranging between 1 and 10. 
If $C$ is small, statistical fluctuations are more visible, 
and as $C$ increases, the computational time and memory usage become larger. 
Without significant restriction, we stick to $C=5$ in most cases. 

In principle, such a system (Eq.~(\ref{eqn:toysystem})) can easily be extended to larger dimensions by replacing the operators by random 
Kronecker products of $\{\mathbb{I},\frac{1}{2}\sigma_x,\frac{1}{2}\sigma_y,\frac{1}{2}\sigma_z\}$, 
which imitates spin interactions. 
However, we find that associating the resulting eigenvalues and eigenvectors of the evolution operators of different versions 
to each other becomes much more complicated because eigenvalues often become degenerate. 

Thus, for higher dimensions $d>2$, we additionally define a completely random Hamiltonian (Eq.~(\ref{eqn:Hamiltonian})) 
without physical interpretation.

\subsection{Error functions} 

We aim to define a measure for the error of the evolution operator 
\begin{equation}
    U^{[n]}_i(t_0+T,t_0) \,,\, i=1,2,3, 
\end{equation}
with a separate view on the eigenvalues and on the eigenframe. 
The index $i$ indicates the theoretical framework: 
1) AHT, 
2) FVV, 
3) FVV without kick operator. 
We formulate the eigenvalue equation of the unitary operator 
\begin{equation}
        U_i(t_0+T,t_0)\vec{v}_{i,j} = \kappa_{i,j} \vec{v}_{i,j} \,,\quad
        \lvert \kappa_{i,j} \rvert = 1\,, \quad
        \vec{v}_{i,j}^\dagger \vec{v}_{i,k} = \delta_{jk}\,, 
\end{equation}
with the Kronecker delta $\delta_{jk}$,
and corresponding to the design of our systems, all eigenvalues are assumed to be non-degenerate. 
The index $n$ is omitted because the definitions stay the same for each order. 
The error of eigenvalues and eigenvectors for version $i$ is calculated via 
\begin{equation}
    \Delta^\text{val}_i := \sum_{j=1}^{d} \frac{\lvert \kappa_{i,j}-\kappa_{\text{true},j}\rvert}{\lvert\kappa_{\text{true},j}\rvert}
    = \sum_{j=1}^{d} {\lvert \kappa_{i,j}-\kappa_{\text{true},j}\rvert} \,,
\end{equation}
\begin{equation}
    \Delta^\text{vec}_i := \sum_{j=1}^{d}  \arccos(\lvert \vec{v}_{\text{true},j}^\dagger \vec{v}_{i,j} \rvert ) \,. 
\end{equation}
The true eigenvalues $\{\kappa_{\text{true},j}\}$ and eigenvectors $\{\vec{v}_{\text{true},j}\}$ belong to a numerically exact evolution operator. 
The definition of $\Delta^\text{vec}_i$ takes into account that the normalized eigenvectors are unique apart from
a phase factor $e^{i\phi}$, $\phi\in\mathbb{R}$. 


\section{Role of the kick operator in NMR spectra}
\label{app:kick_operator}

In the calculation of NMR spectra based on stroboscopic observations with the help of the FVV approach, 
it seems secondary whether the kick operator is included or not. 
This cannot be understood based on the analytical framework which clearly states that only the version with 
kick operator is accurate. 
Thus, we provide a summary of some numerical investigations to shed at least a 
little light on this issue. 

As mentioned in Eq.~(\ref{eq:difference}), 
the difference between the spectrum 
with and without the kick operator in FVV is given by 
\begin{subequations}
    \begin{align}
        \Delta \mathcal{S}(\omega)
        &= \sum_{jk} ( \tilde{\rho}_{jk} \tilde{\mathcal{S}}_{kj} - \rho_{jk} \mathcal{S}_{kj} ) \frac{1}{1 - e^{-\eta \cdot T} e^{i (\omega -\Delta \lambda_{jk})\cdot T}} \,, \\ 
        \tilde{\rho} &:= e^{+i\Lambda(t_0)} \rho e^{-i\Lambda(t_0)}\,,\quad 
        \tilde{\mathcal{S}} := e^{+i\Lambda(t_0)} \mathcal{S} e^{-i\Lambda(t_0)}\,. 
    \end{align}
\end{subequations} 
For the MAS examples presented in Sect.~\ref{sub:MAS}, $\rho=S^x$ and $\mathcal{S}=S^+$ are used. 
We define 
for any operator $S^{x,y,z,\pm}$ and for $M_{ij} := (S^x)_{ij} (S^+)_{ji}$ 
\begin{subequations}
    \begin{align}
        \tilde{S}^\alpha &:= e^{+i\Lambda(t_0)} S^\alpha e^{-i\Lambda(t_0)} \,,\quad 
        \Delta_1^\alpha := \frac{\lvert\lvert \tilde{S}^\alpha - S^\alpha\rvert\rvert}{\lvert\lvert S^\alpha\rvert\rvert} \,,\quad 
        \alpha=x,y,z,\pm\,,\\ 
        \tilde{M}_{ij} &:= (\tilde{S}^x)_{ij} (\tilde{S}^+)_{ji} \,,\quad 
        \Delta_2 := \frac{\lvert\lvert \tilde{M} - M\rvert\rvert}{\lvert\lvert M\rvert\rvert} \,. 
    \end{align}
\end{subequations}
The matrix norm is the spectral norm, 
and the differences $\Delta_1^\alpha$ and $\Delta_2$ are computed for several systems while varying the frequency $\omega_\text{r}$. 
The first three orders of the kick operator are considered, 
but this does not alter the values of $\Delta_{1/2}$. 
Hence, the following observations refer to the series expansion up to the third order, 
i.e., $\Lambda^{[3]}$. 

\begin{figure*}
    \centering
    \includegraphics[width=0.95\textwidth]{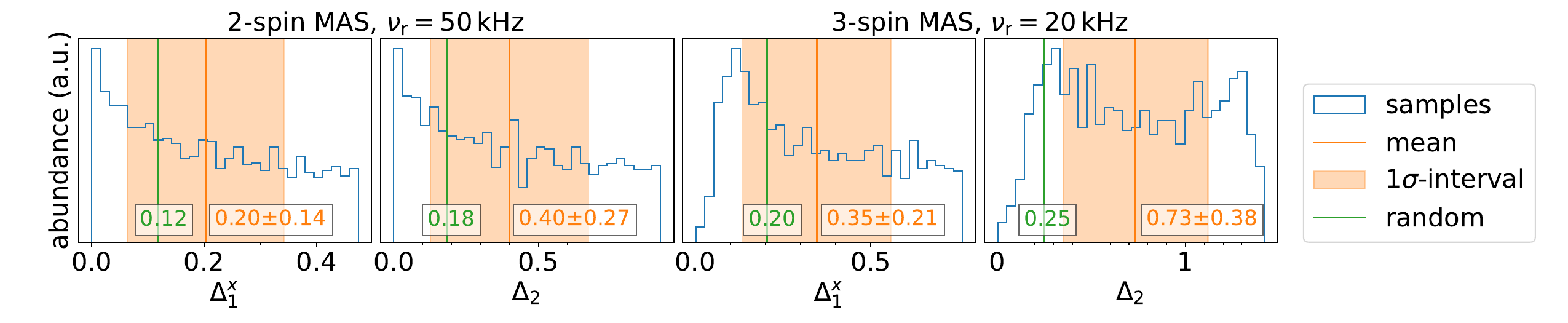}
    \caption{
        The distributions of the relative differences of 
        the operators with and without the kick operator are shown. They
        depend on the crystallite's orientation in the case of MAS.
        The deviation of completely random Hermitian matrices (green) 
        is on average smaller, 
        though their according spectra bear well-observable differences. 
    }
    \label{fig:app_kickoperator}
\end{figure*}

The orientational angles $(\alpha,\beta,\gamma)$ in the MAS NMR Hamiltonian~(\ref{eqn:MASHamiltonian})  
have a large influence on the concrete size of the differences $\Delta_{1/2}$. 
Two histograms for the three-spin system under MAS varying the orientation in Fig.~\ref{fig:app_kickoperator}
visualize their distribution. 
The $S^z$ operator remains unaffected by the kick operator 
in the MAS systems, i.e., $\Delta_1^z=0$. 
This is in contrast to $\Delta_1^{x,y,\pm}$ and $\Delta_2$, 
which are astonishingly consistently smaller for 
entirely random matrices than for MAS Hamiltonians. 
In particular, this means 
$\Delta_{1,\text{random}}^{x,y,\pm} / \Delta_{1,\text{MAS}}^{x,y,\pm}\sim 0.7$
and 
$\Delta_{2,\text{random}} / \Delta_{2,\text{MAS}}\sim 0.4$. 
This is especially surprising given that the spectra of these entirely random 
Hermitian matrices yield well-visible deviations between FVV with and without kick operator. 
Hence, we conclude that the secondary influence of the kick operator on the NMR spectrum is an 
effect of both NMR-specific Hamiltonians and the combined expressions used for the 
calculation of the spectrum in Eq.~(\ref{eqn:spectrum_summary}). 

\bibliographystyle{unsrt} 
\bibliography{references}

@PREAMBLE{
 "\providecommand{\noopsort}[1]{}" 
 # "\providecommand{\singleletter}[1]{#1}%" 
}

@phdthesis{aebischerLineWidthMagicAngle2024,
	title = {Line Width in Magic-Angle Spinning Solid-State {NMR} Spectroscopy},
	doi = {10.3929/ethz-b-000712293},
	school = {{ETH} Zurich},
	type = {Doctoral Thesis},
	author = {Aebischer, Kathrin},
	year = {2024},
}

@BOOK{AbramowitzSegun,
   author = "Abramowitz, Milton and Segun, Irene A.",
   title = "Handbook of Mathematical Functions",
   publisher = "Dover Publications, Inc., New York",
   edition = "9th",
   month = "1~" # jun,
   year = "1965",
}

@article{andrewNuclearMagneticResonance1958,
	title = {Nuclear Magnetic Resonance Spectra from a Crystal rotated at High Speed},
	volume = {182},
	issn = {1476-4687},
	doi = {10.1038/1821659a0},
	pages = {1659--1659},
	number = {4650},
	journal = {Nat.},
	author = {Andrew, E. R. and Bradbury, A. and Eades, R. G.},
	year = {1958},
}

@article{andrewRemovalDipolarBroadening1959,
	title = {Removal of Dipolar Broadening of Nuclear Magnetic Resonance Spectra of Solids by Specimen Rotation},
	volume = {183},
	issn = {1476-4687},
	doi = {10.1038/1831802a0},
	pages = {1802--1803},
	number = {4678},
	journal = {Nat.},
	author = {Andrew, E. R. and Bradbury, A. and Eades, R. G.},
	year = {1959},
}

@article{arnalExponentialPerturbativeExpansions2020,
	title = {Exponential Perturbative Expansions and Coordinate Transformations},
	volume = {25},
	issn = {2297-8747},
	doi = {10.3390/mca25030050},
	pages = {50},
	number = {3},
	journal = {Math. comput. appl.},
	author = {Arnal, Ana and Casas, Fernando and Chiralt, Cristina},
	year = {2020},
	langid = {english},
}

@article{BaldusLevanteMeier+1994+80+88,
title = {Numerical Simulation of Magnetic Resonance Experiments: Concepts and Applications to Static, Rotating and Double Rotating Experiments},
author = {M. Baldus and T. O. Levante and B. H. Meier},
pages = {80--88},
volume = {49},
number = {1-2},
journal = {Z. Naturforsch. A.},
doi = {10.1515/zna-1994-1-214},
year = {1994},
}

@article{blanesMagnusExpansionIts2009,
	title = {The Magnus expansion and some of its applications},
	volume = {470},
	issn = {0370-1573},
	doi = {10.1016/j.physrep.2008.11.001},
	pages = {151--238},
	number = {5},
	journal = {Phys. Rep.},
	author = {Blanes, S. and Casas, F. and Oteo, J. A. and Ros, J.},
	year = {2009},
}

@article{blochMagneticResonanceforNonrotatingField1940,
	title = {Magnetic Resonance for Nonrotating Fields},
	volume = {57},
	issn = {0031-899X},
	doi = {10.1103/PhysRev.57.522},
	pages = {522--527},
	number = {6},
	journal = {Phys. Rev.},
	author = {Bloch, F. and Siegert, A.},
	year = {1940},
}

@book{BrinkSatchler1993,
	location = {Oxford, New York},
	title = {Angular Momentum},
	publisher = {Clarendon Press},
	author = {Brink, David M. and Satchler, George R.},
	year = {1993},
}

@article{Brinkmann2004,
    author = {Brinkmann, Andreas and Edén, Mattias},
    title = {Second order average Hamiltonian theory of symmetry-based pulse schemes in the nuclear magnetic resonance of rotating solids: Application to triple-quantum dipolar recoupling},
    journal = {J. Chem. Phys.},
    volume = {120},
    number = {24},
    pages = {11726-11745},
    year = {2004},
    issn = {0021-9606},
    doi = {10.1063/1.1738102},
}

@article{bukovUniversalHighFrequencyBehavior2015,
	title = {Universal high-frequency behavior of periodically driven systems: from dynamical stabilization to Floquet engineering},
	volume = {64},
	issn = {0001-8732, 1460-6976},
	doi = {10.1080/00018732.2015.1055918},
	pages = {139--226},
	number = {2},
	journal = {Adv. Phys.},
	author = {Bukov, Marin and D'Alessio, Luca and Polkovnikov, Anatoli},
	year = {2015},
}

@article{casasFloquetTheoryExponential2001,
	title = {Floquet theory: exponential perturbative treatment},
	volume = {34},
	issn = {0305-4470},
	doi = {10.1088/0305-4470/34/16/305},
	pages = {3379},
	number = {16},
	journal = {J. Phys. A: Math. Gen.},
	author = {Casas, F. and Oteo, J. A. and Ros, J.},
	year = {2001},
}

@article{chengInvestigationsNonrandomNumerical1973,
	title = {Investigations of a nonrandom numerical method for multidimensional integration},
	volume = {59},
	issn = {0021-9606},
	doi = {10.1063/1.1680590},
	pages = {3992--3999},
	number = {8},
	journal = {J. Chem. Phys.},
	author = {Cheng, Vera B. and Suzukawa, Jr., Henry H. and Wolfsberg, Max},
	year = {1973},
}

@article{Dengis2025,
  title = {Accelerated creation of NOON states with ultracold atoms via counterdiabatic driving},
  author = {Dengis, Simon and Wimberger, Sandro and Schlagheck, Peter},
  journal = {Phys. Rev. A},
  volume = {111},
  issue = {3},
  pages = {L031301},
  numpages = {6},
  year = {2025},
  month = {Mar},
  publisher = {American Physical Society},
  doi = {10.1103/PhysRevA.111.L031301},
}

@article{Dyson1949,
  title = {The $S$ Matrix in Quantum Electrodynamics},
  author = {Dyson, F. J.},
  journal = {Phys. Rev.},
  volume = {75},
  issue = {11},
  pages = {1736--1755},
  numpages = {0},
  year = {1949},
  publisher = {American Physical Society},
  doi = {10.1103/PhysRev.75.1736},
}

@article{eckardtHighfrequencyApproximationPeriodically2015,
	title = {High-frequency approximation for periodically driven quantum systems from a Floquet-space perspective},
	volume = {17},
	issn = {1367-2630},
	doi = {10.1088/1367-2630/17/9/093039},
	pages = {093039},
	number = {9},
	journal = {New J. Phys.},
	publisher = {{IOP} Publishing},
	author = {Eckardt, André and Anisimovas, Egidijus},
	year = {2015},
}

@article{eckardtSuperfluidInsulatorTransitionPeriodically2005,
	title = {Superfluid-Insulator Transition in a Periodically Driven Optical Lattice},
	volume = {95},
	issn = {0031-9007, 1079-7114},
	doi = {10.1103/PhysRevLett.95.260404},
	pages = {260404},
	number = {26},
	journal = {Phys. Rev. Lett.},
	author = {Eckardt, André and Weiss, Christoph and Holthaus, Martin},
	year = {2005},
}

@article{floquetEquationsDifferentiellesLineaires1883,
	title = {Sur les équations différentielles linéaires à coefficients périodiques},
	volume = {12},
	issn = {1873-2151},
	doi = {10.24033/asens.220},
	pages = {47--88},
	journal = {Ann. Sci. Éc. Norm. Supér.},
	author = {Floquet, G.},
	year = {1883},
}

@article{haeberlenCoherentAveragingEffects1968,
	title = {Coherent Averaging Effects in Magnetic Resonance},
	volume = {175},
	doi = {10.1103/PhysRev.175.453},
	pages = {453--467},
	number = {2},
	journal = {Phys. Rev.},
	author = {Haeberlen, U. and Waugh, J. S.},
	year = {1968},
}

@book{haeberlenHighResolutionNMR2012,
	title = {High Resolution {NMR} in Solids Selective Averaging},
	isbn = {978-0-12-025561-0},
	author = {Haeberlen, Ulrich},
    publisher = {Academic Press}, 
	year = {1976},
}

@article{Hou2012,
    author = {Hou, Guangjin and Byeon, In-Ja L. and Ahn, Jinwoo and Gronenborn, Angela M. and Polenova, Tatyana},
    title = {Recoupling of chemical shift anisotropy by R-symmetry sequences in magic angle spinning NMR spectroscopy},
    journal = {J. Chem. Phys.},
    volume = {137},
    number = {13},
    pages = {134201},
    year = {2012},
    month = {10},
    issn = {0021-9606},
    doi = {10.1063/1.4754149},
}

@article{IVANOV202117,
title = {Floquet theory in magnetic resonance: Formalism and applications},
journal = {Prog. Nucl. Magn. Reson. Spectrosc.},
volume = {126-127},
pages = {17-58},
year = {2021},
issn = {0079-6565},
doi = {10.1016/j.pnmrs.2021.05.002},
author = {Konstantin L. Ivanov and Kaustubh R. Mote and Matthias Ernst and Asif Equbal and Perunthiruthy K. Madhu},
}

@article{leskesFloquetTheorySolidstate2010a,
	title = {Floquet theory in solid-state nuclear magnetic resonance},
	volume = {57},
	issn = {0079-6565},
	doi = {10.1016/j.pnmrs.2010.06.002},
	pages = {345--380},
	number = {4},
	journal = {Prog. Nucl. Magn. Reson. Spectrosc.},
	author = {Leskes, Michal and Madhu, P. K. and Vega, Shimon},
	year = {2010},
}

@article{Levante10121995,
author = {T.O. Levante and M. Baldus and B.H. Meier and R.R. Ernst},
title = {Formalized quantum mechanical Floquet theory and its application to sample spinning in nuclear magnetic resonance},
journal = {Mol. Phys.},
volume = {86},
number = {5},
pages = {1195--1212},
year = {1995},
publisher = {Taylor \& Francis},
doi = {10.1080/00268979500102671}
}

@article{loweFreeInductionDecays1959,
	title = {Free Induction Decays of Rotating Solids},
	volume = {2},
	doi = {10.1103/PhysRevLett.2.285},
	pages = {285--287},
	number = {7},
	journal = {Phys. Rev. Lett.},
	author = {Lowe, I. J.},
	year = {1959},
}

@article{magnus1954,
	author = {Magnus, Wilhelm},
	title = {On the exponential solution of differential equations for a linear operator},
	journal = {Commun. Pure Appl. Math.},
	volume = {7},
	number = {4},
	pages = {649-673},
	doi = {10.1002/cpa.3160070404},
	year = {1954}
}

@article{manangaIntroductionFloquetMagnusExpansion2011,
	title = {Introduction of the Floquet-Magnus expansion in solid-state nuclear magnetic resonance spectroscopy},
	volume = {135},
	issn = {1089-7690},
	doi = {10.1063/1.3610943},
	pages = {044109},
	number = {4},
	journal = {J. Chem. Phys.},
	author = {Mananga, Eugène S. and Charpentier, Thibault},
	year = {2011},
}

@article{mikamiBrillouinWignerTheoryHighfrequency2016,
	title = {Brillouin-Wigner theory for high-frequency expansion in periodically driven systems: Application to Floquet topological insulators},
	volume = {93},
	issn = {2469-9950, 2469-9969},
	doi = {10.1103/PhysRevB.93.144307},
	pages = {144307},
	number = {14},
	journal = {Phys. Rev. B},
	author = {Mikami, Takahiro and Kitamura, Sota and Yasuda, Kenji and Tsuji, Naoto and Oka, Takashi and Aoki, Hideo},
	year = {2016},
	eprinttype = {arxiv},
}

@article{oka2019,
   author = "Oka, Takashi and Kitamura, Sota",
   title = "Floquet Engineering of Quantum Materials", 
   journal= "Annu. Rev. Condens. Matter Phys.",
   year = "2019",
   volume = "10",
   pages = "387-408",
   doi = "https://doi.org/10.1146/annurev-conmatphys-031218-013423",
   issn = "1947-5462",
}

@article{oonAverageHamiltonianTheory2026,
	title = {Beyond average Hamiltonian theory for quantum sensing},
	volume = {8},
	doi = {10.1103/w2sf-zxdk},
	pages = {013222},
	number = {1},
	journal = {Phys. Rev. Res.},
	publisher = {American Physical Society},
	author = {Oon, Jner Tzern and Carrasco, Sebastian C. and Hart, Connor A. and Witt, George A. and Malinovsky, Vladimir S. and Walsworth, Ronald},
	year = {2026},
}

@article{pauliZurQuantenmechanikMagnetischen1927,
	title = {Zur Quantenmechanik des magnetischen Elektrons},
	volume = {43},
	issn = {0044-3328},
	doi = {10.1007/BF01397326},
	pages = {601--623},
	number = {9},
	journal = {Zeitschrift für Physik},
	author = {Pauli, W.},
	year = {1927},
}

@article{PILEIO200765,
title = {Analytical theory of $\gamma$-encoded double-quantum recoupling sequences in solid-state nuclear magnetic resonance},
journal = {J. Magn. Reson.},
volume = {186},
number = {1},
pages = {65-74},
year = {2007},
issn = {1090-7807},
doi = {10.1016/j.jmr.2007.01.009},
author = {Giuseppe Pileio and Maria Concistrè and Neville McLean and Axel Gansmüller and Richard C.D. Brown and Malcolm H. Levitt},
}

@article{SchmidtVega1992,
    author = {Schmidt, Asher and Vega, Shimon},
    title = {The Floquet theory of nuclear magnetic resonance spectroscopy of single spins and dipolar coupled spin pairs in rotating solids},
    journal = {J. Chem. Phys.},
    volume = {96},
    number = {4},
    pages = {2655-2680},
    year = {1992},
    month = {02},
    issn = {0021-9606},
    doi = {10.1063/1.462015},
}

@article{scholzOperatorbasedFloquetTheory2010a,
	title = {Operator-based Floquet theory in solid-state {NMR}},
	volume = {37},
	issn = {0926-2040},
	doi = {10.1016/j.ssnmr.2010.04.003},
	pages = {39--59},
	number = {3},
	journal = {Solid State Nucl. Magn. Reson.},
	author = {Scholz, Ingo and van Beek, Jacco D. and Ernst, Matthias},
	year = {2010},
}

@article{vanvlecksigmaTypeDoublingElectron1929,
	title = {On $\sigma$-Type Doubling and Electron Spin in the Spectra of Diatomic Molecules},
	volume = {33},
	doi = {10.1103/PhysRev.33.467},
	pages = {467--506},
	number = {4},
	journal = {Phys. Rev.},
	author = {Van Vleck, J. H.},
	year = {1929},
}

@incollection{Vega1996,
   author    = "Vega, S.",
   title     = "Floquet theory",
   pages     = "2011-2025",
   editor    = "Grant, D.M. and Harry, R.K.",
   booktitle = "Encyclopedia of Nuclear Magnetic Resonance",
   publisher = "Wiley",
   address   = "New York",
   year      = "1996",
}

@article{wilcoxExponentialOperatorsParameter1967,
	title = {Exponential Operators and Parameter Differentiation in Quantum Physics},
	volume = {8},
	issn = {0022-2488, 1089-7658},
	doi = {10.1063/1.1705306},
	pages = {962--982},
	number = {4},
	journal = {J. Math. Phys.},
	author = {Wilcox, R. M.},
	year = {1967},
}

@article{EVANS196872,
	title = {On some applications of the Magnus expansion in nuclear magnetic resonance},
	journal = {Ann. Phys. (N. Y.)},
	volume = {48},
	number = {1},
	pages = {72-93},
	year = {1968},
	issn = {0003-4916},
	doi = {10.1016/0003-4916(68)90270-4},
	author = {W. A. B. Evans},
}

@article{Goldman2014,
  title = {Periodically Driven Quantum Systems: Effective Hamiltonians and Engineered Gauge Fields},
  author = {Goldman, N. and Dalibard, J.},
  journal = {Phys. Rev. X},
  volume = {4},
  issue = {3},
  pages = {031027},
  numpages = {29},
  year = {2014},
  month = {Aug},
  publisher = {American Physical Society},
  doi = {10.1103/PhysRevX.4.031027},
}

@article{Haeberlen1968,
  title = {Coherent Averaging Effects in Magnetic Resonance},
  author = {Haeberlen, U. and Waugh, J. S.},
  journal = {Phys. Rev.},
  volume = {175},
  issue = {2},
  pages = {453--467},
  year = {1968},
  month = {Nov},
  doi = {10.1103/PhysRev.175.453},
}

@book{mehringPrinciplesHighResolution2012,
	edition = {2},
	title = {Principles of High Resolution NMR in Solids},
	isbn = {978-0-12-025561-0},
	author = {Mehring, Michael},
    publisher = {Springer}, 
	year = {1983},
}

@article{brinkmannIntroductionAverageHamiltonian2016a,
	title = {Introduction to average Hamiltonian theory. I. Basics},
	volume = {45A},
	issn = {1546-6086, 1552-5023},
	doi = {10.1002/cmr.a.21414},
	pages = {e21414},
	number = {6},
	journal = {Concepts Magn. Reson. Part A},
	author = {Brinkmann, Andreas},
	year = {2016},
}

@article{shirleySolutionSchrodingerEquation1965,
	title = {Solution of the Schrödinger Equation with a Hamiltonian Periodic in Time},
	volume = {138},
	doi = {10.1103/PhysRev.138.B979},
	pages = {B979--B987},
	number = {4},
	journal = {Phys. Rev.},
	author = {Shirley, Jon H.},
	year = {1965},
}

@article{Zhou2020,
  title = {Quantum Metrology with Strongly Interacting Spin Systems},
  author = {Zhou, Hengyun and Choi, Joonhee and Choi, Soonwon and Landig, Renate and Douglas, Alexander M. and Isoya, Junichi and Jelezko, Fedor and Onoda, Shinobu and Sumiya, Hitoshi and Cappellaro, Paola and Knowles, Helena S. and Park, Hongkun and Lukin, Mikhail D.},
  journal = {Phys. Rev. X},
  volume = {10},
  issue = {3},
  pages = {031003},
  numpages = {9},
  year = {2020},
  month = {Jul},
  publisher = {American Physical Society},
  doi = {10.1103/PhysRevX.10.031003},
}

@article{Ponti:1999td, 
year = {1999}, 
title = {{Simulation of magnetic resonance static powder lineshapes: A quantitative assessment of spherical codes}}, 
author = {Ponti, A}, 
journal = {Journal of Magnetic Resonance}, 
issn = {1090-7807}, 
doi = {10.1006/jmre.1999.1758}, 
pmid = {10341133}, 
url = {http://links.isiglobalnet2.com/gateway/Gateway.cgi?GWVersion=2\&SrcAuth=mekentosj\&SrcApp=Papers\&DestLinkType=FullRecord\&DestApp=WOS\&KeyUT=000080676000012}, 
pages = {288 -- 297}, 
number = {2}, 
volume = {138}, 
language = {English}, 
month = {00}
}

\end{document}